\documentclass{lmcs}
\pdfoutput=1
\usepackage[utf8]{inputenc}

\usepackage{lastpage}
\lmcsdoi{17}{4}{3}
\lmcsheading{}{\pageref{LastPage}}{}{}%
{Jan.~26,~2021}{Oct.~25,~2021}{}

\pdfoutput=1
\theoremstyle{plain}\newtheorem{sta}[thm]{Statement} %
\usepackage{amssymb}
\usepackage{setspace,xspace}
\usepackage{booktabs}
\usepackage{prooftree}
\newcommand{\NotIn}{\not\in}
\newcommand{\utA}{\tspc \boldsymbol{t}_A \tspc}
\newcommand{\tA}{\tspc t_A \tspc}
\newcommand{\dtLBrack}{$\lbrack\tspc$\xspace}
\newcommand{\dtRBrack}{$\tspc\rbrack$\xspace}
\newcommand{\BS}[1]{\tspc \boldsymbol{#1} \tspc}
\newcommand{\CSpFive}{\mathsf{C}^{\mathrm{\prime}}_{\mathrm{S5}}}
\newcommand{\PcsFive}{\mathsf{C}_{\mathrm{S5}}}
\newcommand{\Minlog}{\texttt{Minlog}\xspace}
\newcommand{\dtRho}{\tspc \rho \tspc}
\newcommand{\dtSgm}{\tspc \sigma \tspc}
\newcommand{\dtTau}{\tspc \tau \tspc}

\newcommand{\AZuY}{\tspc A \tspc ( \tspc Z \uY \tspc ) \tspc}
\newcommand{\AZuYs}{\tspc A \tspc ( \tspc Z \uY ) \tspc}

\newcommand{\ZtCmUh}{\tspc Z, \uH}
\newcommand{\uttpv}{\ut \tspc \utp \uv}
\newcommand{\utpN}{ \utp [ \tspc n \tspc ] }
\newcommand{\utpSucN}{ \utp [ \Succ n \tspc ] }
\newcommand{\utpZero}{ \utp [ \tspc \Zero \tspc ] }
\newcommand{\uztSucN}{ \uzt [ \Succ n \tspc ] }
\newcommand{\uztN}{ \uzt [ \tspc n \tspc ] }
\newcommand{\uztGamZ}{ \uztGama [ \tspc \Zero \tspc ] }
\newcommand{\uztSnUV}{\uzt \nspc [ \nspc \Succ \tspc n \nspc ] \nspc \uv}
\newcommand{\uztsSnUV}{\uzts \nspc [ \nspc \Succ \tspc n \nspc ] \nspc \uv}
\newcommand{\uztsNuttpv}{\uzts \tspc [ \tspc n \tspc ] \dspc ( \uttpv )}
\newcommand{\UdtpUtpUv}{\udtp \tspc [ \tspc \utp ; \tspc \uv \tspc ]}
\newcommand{\nmvb}{\mathnormal{|}}
\newcommand{\AndSlashOr}{and\hspace{.16em}/\hspace{.16em}or\xspace}
\newcommand{\ul}[1]{\underline{#1}}
\newcommand{\ol}[1]{ \overline{ #1 } }
\newcommand{\tld}[1]{ \widetilde{ #1 } }

\newcommand{\converts}{\hookrightarrow}
\newcommand{\tspc}{ \; \! }
\newcommand{\mssp}{}
\newcommand{\nspc}{ \, }
\newcommand{\dtcsp}{ \tspc , \tspc }
\newcommand{\dspc}{ \ \ }
\newcommand{\qqquad}{ \qquad \quad }

\newcommand{\pCite}[1]{\cite{#1}\mssp.}
\newcommand{\scCite}[1]{\cite{#1}\mssp;}
\newcommand{\cmCite}[1]{\cite{#1}\mssp,}
\newcommand{\lpgref}[1]{\mssp\pageref{#1}\mssp}
\newcommand{\lref}[1]{\mssp\ref{#1}\mssp}
\newcommand{\ptref}[1]{\mssp(\ref{#1})\mssp}
\newcommand{\reffl}[1]{\mssp(\mssp\ref{#1}\mssp)\mssp\xspace}
\newcommand{\reffL}[1]{\,(\mssp\ref{#1}\mssp)\,\xspace}
\newcommand{\refc}[1]{\ref{#1}\mssp,}
\newcommand{\reffc}[1]{\mssp\ptref{#1}\mssp,}
\newcommand{\reffLc}[1]{$\;$\ptref{#1}\mssp,}
\newcommand{\refp}[1]{\ref{#1}\mssp.}
\newcommand{\reffp}[1]{\mssp\ptref{#1}\mssp.}
\newcommand{\reffLp}[1]{$\;$\ptref{#1}\mssp.}
\newcommand{\ND}{\mbox{ND}\xspace}
\newcommand{\lND}{ Natural Deduction\xspace}
\newcommand{\FVC}{\mbox{\rm(c)}\xspace}

\newcommand{\ie}{i.e.,\mssp\xspace}
\newcommand{\eg}{e.g.,\mssp\xspace}
\newcommand{\Eeg}{E.g.,\mssp\xspace}
\newcommand{\Tpymphc}{$\Typ$-$\tspc$polymorphic\xspace}
\newcommand{\Tunrealizable}{$\Term\!$-$\tspc$unrealizable\xspace}
\newcommand{\Tunrealizability}{$\Term\!\!-\tspc$unrealizability\xspace}
\newcommand{\SQT}[1]{`#1'}
\newcommand{\QT}[1]{``#1''}

\newcommand{\sfFV}{ \sf FV }
\newcommand{\FV}[1]{ {\sfFV} \tspc ( \tspc #1 \tspc ) }
\newcommand{\Array}[2]{\[\begin{array}{#1}#2\end{array}\]}
\newcommand{\mArray}[2]{\(\begin{array}{#1}#2\end{array}\)}
\newcommand{\inArray}[2]{\begin{array}{#1}#2\end{array}}
\newcommand{\lbmh}[1]{$\tspc #1 $\xspace}
\newcommand{\llbmh}[1]{\mbox{\( \nspc #1 \nspc \)}}
\newcommand{\Lbmh}[1]{\mbox{\( \; #1 \; \)}}
\newcommand{\bmh}[1]{\mbox{\( #1 \)}}

\newcommand{\bmsc}[1]{\mbox{$ #1 \tspc $;}}

\newcommand{\bmp}[1]{\mbox{$ #1 \tspc $.}}
\newcommand{\lbmp}[1]{\mbox{$ \, #1 \tspc$.}}
\newcommand{\Lbmp}[1]{\mbox{$ \ #1 \tspc$.}}
\newcommand{\lbmc}[1]{\mbox{$ \, #1 \tspc$,}}
\newcommand{\Lbmc}[1]{\mbox{$ \ #1 \tspc$,}}
\newcommand{\bmc}[1]{\mbox{$ #1 \tspc$,}}
\newcommand{\bmcc}[1]{\mbox{$ #1 \tspc$:}}
\newcommand{\NN}{ \tspc \mathbb{N} \tspc }
\newcommand{\BB}{ \tspc \mathbb{B} \tspc }
\newcommand{\lbda}[2]{\nspc \mathnormal{\lambda \tspc #1 \tspc .\ #2} \nspc}
\newcommand{\set}[1]{ \left\{ \tspc #1 \tspc \right\} }
\newcommand{\bnor}{\, \nmvb \,\,}
\newcommand{\bndef}{\mathtt{::=}\ }
\newcommand{\pair}[2]{\langle \tspc #1 \tspc / \tspc #2 \tspc \rangle}

\newcommand{\withparam}[2]{ #1_{\!\lbrack{#2}\rbrack} }
\newcommand{\mtparam}[2]{$ \lbrack\ #1, \, #2\ \rbrack $ }
\newcommand{\Typ}{\tspc \mathnormal{T} \tspc}
\newcommand{\Term}{\tspc \mathcal{T} \tspc}

\newcommand{\EFQ}{\mathtt{EFQ}}
\newcommand{\Stab}{\mathtt{Stab}}

\newcommand{\pt}{\prooftree}
\newcommand{\ept}{\endprooftree}
\newcommand{\jst}{\vspace{1pt} \justifies \vspace{1pt}}
\newcommand{\usg}{\, \using \ }

\newcommand{\BvdashS}{\BS{\vdash}}
\newcommand{\prfg}{\,{\BvdashS}\,}
\newcommand{\DB}[1]{\raisebox{-2pt}{$#1$}}
\newcommand{\lprg}{\,{\BvdashS}\raisebox{-0.6ex}{\scriptsize $\!\!\BS{l}$}\ }
\newcommand{\mprfg}{\,{\BvdashS}\raisebox{+1.2ex}{\scriptsize $\!\!\!\!\!\BS{m}$}\,}
\newcommand{\mlprg}{\,{\BvdashS}\raisebox{+1.2ex}{\scriptsize $\!\!\!\!\!\BS{m}$}\raisebox{-0.7ex}{\scriptsize $\!\!\!\!\!\BS{l}$}\ }

\newcommand{\mnEEqBD}{\mathnormal{:\equiv}}
\newcommand{\EqBD}{\nspc \mnEEqBD \tspc}
\newcommand{\EqBDl}{\, \mnEEqBD \,}
\newcommand{\EqBDL}{\ \mnEEqBD \ }
\newcommand{\mnEQuiv}{\mathnormal{\equiv}}
\newcommand{\tEquiv}{\tspc \mnEQuiv \tspc}
\newcommand{\Equiv}{\nspc \mnEQuiv \tspc}
\newcommand{\EquiVl}{\; \mnEQuiv \;}
\newcommand{\EquIVll}{\, \mnEQuiv \ }
\newcommand{\EquivL}{\ \mnEQuiv \ }
\newcommand{\mnTrue}{\mathnormal{\mathtt{T}}}
\newcommand{\mnFalse}{\mathnormal{\mathtt{F}}}
\newcommand{\BolTru}{\tspc \mnTrue \tspc}
\newcommand{\BlFals}{\tspc \mnFalse \tspc}
\newcommand{\mnBot}{\mathnormal{\bot}}
\newcommand{\mnTop}{\mathnormal{\top}}
\newcommand{\LogTruth}{\tspc \mnTop \tspc}
\newcommand{\LogFalsity}{\tspc \mnBot \tspc}
\newcommand{\impnc}{\texttt{-->}\xspace}
\newcommand{\mnLImp}{\mathnormal{\to}}
\newcommand{\dtimp}{\; \mnLImp \;}
\newcommand{\kimp}{\; \mathnormal{\rightarrow_{\mathtt{k}}} \,}
\newcommand{\mnLogQuiv}{\mathnormal{\leftrightarrow}}
\newcommand{\loquiv}{\tspc \mnLogQuiv \tspc}
\newcommand{\dtLoQuiv}{\nspc \mnLogQuiv \nspc}
\newcommand{\mnLNot}{\mathnormal{\lnot}}
\newcommand{\dtnot}{\tspc \mnLNot \tspc}
\newcommand{\notK}{\tspc \mnLNot_{\mathtt{k}} \tspc}
\newcommand{\twonot}{\tspc \mnLNot \tspc \mnLNot \tspc}
\newcommand{\mnLAnd}{\mathnormal{\land}}
\newcommand{\dtand}{\; \mnLAnd \;}
\newcommand{\cLand}{\, \tilde{\mnLAnd} \,}
\newcommand{\mnBox}{\mathnormal{\Box}}
\newcommand{\necsy}{\nspc \mnBox \tspc}
\newcommand{\necsN}{\nspc \mnBox \nspc}
\newcommand{\mnBang}{\mathnormal{!}}
\newcommand{\lnecsy}{\tspc \mnBang \; \!}
\newcommand{\kNecsy}{\tspc \mnBang_{\mathtt{k}} \nspc}
\newcommand{\mnDiamond}{\mathnormal{\Diamond}}
\newcommand{\possy}{\nspc \mnDiamond \tspc}
\newcommand{\wpossy}{\nspc
\raisebox{2pt}{$\tld{\raisebox{-2pt}{$\mnDiamond$}}$} \tspc}
\newcommand{\cpossy}{\nspc \mnLNot \tspc \necsy \tspc \mnLNot \nspc}

\newcommand{\plh}{\diamond}

\newcommand{\etup}{\sqcup}
\newcommand{\nothing}{\sqcup}
\newcommand{\eset}{ \tspc \emptyset \tspc }
\newcommand{\yset}{ \emptyset }
\newcommand{\sset}{ \tspc \mathnormal{\subset} \tspc }
\newcommand{\sseq}{ \tspc \mathnormal{\subseteq} \tspc }

\newcommand{\prm}[1]{ {#1}^{\prime} }
\newcommand{\scd}[1]{ {#1}^{\prime\prime} }

\newcommand{\RgRes}{\mbox{($\pm$)}\xspace}
\newcommand{\PlRes}{\mbox{($+$)}\xspace}
\newcommand{\cmPlRes}{\mbox{($+$)\mssp,}\xspace}
\newcommand{\MnRes}{\mbox{($-$)}\xspace}
\newcommand{\NCRes}{\mbox{($\eset$)}\xspace}
\newcommand{\avar}[2]{\tspc #1 \!: \! #2 \tspc}

\newcommand{\spDelta}{\tspc \Delta \tspc}
\newcommand{\spGamma}{\tspc \Gamma \tspc}
\newcommand{\Gmnsama}{\tspc \refir\Gamma \tspc}
\newcommand{\If}{\tspc \mathtt{If} \tspc}
\newcommand{\Eq}{\tspc \mathtt{Eq} \tspc}
\newcommand{\EqNatt}{\tspc \mathnormal{\mathtt{Eq}_{\tspc\NN}} \tspc}
\newcommand{\EqBool}{\tspc \mathnormal{\mathtt{Eq}_{\tspc\BB}} \tspc}

\newcommand{\atom}[1]{\tspc \mathnormal{ \mathtt{at}\tspc ( \tspc #1 \tspc ) } \tspc}

\newcommand{\Odd}[1]{\tspc \mathnormal{\mathtt{Odd}(#1)} \tspc}
\newcommand{\id}{ \mathtt{(id)} }
\newcommand{\LimpI}{\tspc \mathtt{\rightarrow^{\!i}} \tspc }
\newcommand{\LimpE}{\tspc \mathtt{\rightarrow^{\!e}} \tspc }
\newcommand{\landI}{\tspc \mathtt{\land^{\!i}} \tspc }
\newcommand{\landEl}{\tspc \mathtt{\land^{\!e}_0} \tspc }
\newcommand{\landEr}{\tspc \mathtt{\land^{\!e}_1} \tspc }
\newcommand{\intro}[1]{\mathtt{#1^i}}
\newcommand{\elim}[1]{\mathtt{#1^e}}
\newcommand{\mINcsy}{\mnBox^{\mbox{\footnotesize$\mathrm{i}$}}}
\newcommand{\InecsYsp}{$\tspc\mnBox^{\mbox{\footnotesize$\mathrm{i}$}}\tspc$}
\newcommand{\InecsY}{$\mnBox^{\mbox{\footnotesize$\mathrm{i}$}}$\xspace}
\newcommand{\InecsYb}{$\mnBox^{\mbox{\footnotesize$\mathrm{i}$}}$}
\newcommand{\Bfa}{  \tspc \forall }
\newcommand{\fa}{ \Bfa \tspc }
\newcommand{\FaPlh}{ \Bfa_{ \mbox{\footnotesize$\plh$} } }
\newcommand{\faIB}{ \intro \Bfa }
\newcommand{\faEB}{ \elim \Bfa }

\newcommand{\faE}[1]{ \withparam{ \faEB }{ #1 } }
\newcommand{\Bfanc}{ \forall_{\!\yset} }
\newcommand{\fancIB}{ \tspc \intro \Bfanc }

\newcommand{\fanc}{ \Bfanc \tspc }
\newcommand{\Bfapl}{ \forall_{\!+} }
\newcommand{\faplIB}{ \tspc \intro \Bfapl }

\newcommand{\fapl}{ \Bfapl \tspc }
\newcommand{\Bfamn}{ \forall_{\!-} }
\newcommand{\famnIB}{ \tspc \intro \Bfamn}

\newcommand{\famn}{ \Bfamn \tspc }
\newcommand{\Bfaplh}{ \forall_{\!\plh} }
\newcommand{\faplhIB}{ \tspc \intro \Bfaplh}
\newcommand{\faplhEB}{ \tspc \elim \Bfaplh}
\newcommand{\faplh}{ \Bfaplh \tspc }
\newcommand{\iex}{ \tspc \exists \tspc }
\newcommand{\iexnc}{ \tspc \iex_\yset \tspc }
\newcommand{\ex}{ \tspc \tld{\exists} \tspc}
\newcommand{\ExPlh}{ \tld{\exists}_{ \mbox{\footnotesize$\plh$} } }
\newcommand{\exnc}{ \tspc \ex_\yset \tspc}
\newcommand{\expl}{ \tspc \ex_+ \tspc}
\newcommand{\exmn}{ \tspc \ex_- \tspc}
\newcommand{\exf}{ \tspc \ex \tspc}
 
\newcommand{\LCnRlTx}{\mbox{$\mathcal C_l$}\xspace}
\newcommand{\CnRlTx}{\mbox{$\sf C$}\xspace }
\newcommand{\mLCnRl}{\tspc \mathcal C_l \tspc}
\newcommand{\modCnRl}{\tspc \mathcal C_m \tspc}
\newcommand{\mCnRl}{\tspc \sf C \tspc}
\newcommand{\metaVSys}{\mathrm{VSys}}
\newcommand{\metaISys}{\mathrm{ISys}}
\newcommand{\Fmla}{ \tspc \mathcal{F} \tspc }
\newcommand{\FmlaL}{ \tspc \mathcal{F}_l \tspc }
\newcommand{\FmlaM}{ \tspc \mathcal{F}^m \tspc }
\newcommand{\FmlaML}{ \tspc \mathcal{F}^m_l \tspc }
\newcommand{\GT}{ \tspc {\sf T} \tspc }
\newcommand{\HAom}{ \tspc {\sf HA}^\omega \tspc }
\newcommand{\PAom}{ \tspc {\sf PA}^\omega \tspc }
\newcommand{\ILom}{ \tspc {\sf IL}^\omega \tspc }
\newcommand{\VSys}{ \tspc {\sf NA} \tspc }
\newcommand{\ISys}{ \tspc {\sf NA_{\mathnormal{l}}} \tspc }
\newcommand{\MSys}{ \tspc {\sf NA^{\mathnormal{m}}} \tspc }
\newcommand{\MSysL}{ \nspc {\sf NA^{\mathnormal{m}}_{\mathnormal{l}}} \nspc }

\newcommand{\ldInt}[3]{\tspc \nmvb \nspc {#1} \nspc
  \nmvb^{\raisebox{2pt}{\small{$\, #2$}}}_{\raisebox{-3pt}{\small{$\, #3$}}}\tspc }
\newcommand{\InLdInt}[3]{\tspc \nmvb \nspc #1 \nspc
  \nmvb^{\raisebox{1pt}{\footnotesize{$\, #2$}}}_{\raisebox{-1pt}{\footnotesize{$\, #3$}}}\tspc }
\newcommand{\ldIntB}[1]{\tspc \nmvb \tspc #1 \tspc \nmvb \tspc}
\newcommand{\cLDint}[3]{\ \lbrack \nspc #1 \nspc
  \rbrack^{\raisebox{3pt}{\small{$\, #2$}}}_{\raisebox{-4pt}{\small{$\, #3$}}}\ }

\newcommand{\ugm}{ \BS{\gamma} }
\newcommand{\ugmp}{ \BS{\prm{\gamma}} }
\newcommand{\udt}{ \BS{\delta} }
\newcommand{\udtp}{ \BS{\prm{\delta}} }
\newcommand{\uro}{ \BS{\rho} }

\newcommand{\uf}{\BS f}
\newcommand{\ug}{\BS g}
\newcommand{\up}{\BS p}
\newcommand{\uH}{\BS H}
\newcommand{\uh}{\BS h}
\newcommand{\uY}{\BS Y}
\newcommand{\uYs}{\uY\!}

\newcommand{\ur}{\BS r}
\newcommand{\us}{\BS s}

\newcommand{\ut}{\BS t}
\newcommand{\tp}{\prm{t}}
\newcommand{\utp}{\BS{\tp}}

\newcommand{\uu}{\BS u}
\newcommand{\uU}{\BS U}
\newcommand{\uv}{\BS v}
\newcommand{\uV}{\BS V}
\newcommand{\uw}{\BS w}
\newcommand{\ux}{\BS x}
\newcommand{\uX}{\BS X}
\newcommand{\uy}{\BS y}
\newcommand{\uz}{\BS z}
\newcommand{\uxp}{\BS{\prm{x}}}
\newcommand{\uyp}{\BS{\prm{y}}}
\newcommand{\uzp}{\BS{\prm{z}}}
\newcommand{\uvp}{\BS{\prm{v}}}

\newcommand{\RecNat}{ \tspc \mathtt{R} \tspc }

\newcommand{\Rec}{ \RecNat }
\newcommand{\Succ}{ \tspc \mathtt{S} \tspc }
\newcommand{\Zero}{ \tspc \mathtt{0} \tspc }

\newcommand{\AxTrue}{ \mathtt{TruAx} }
\newcommand{\AxCompat}{\mathtt{CmpAx}}
\newcommand{\AxCompatM}{\mathtt{CmpAx^m}}
\newcommand{\RlCompat}{\mathtt{CMP}}
\newcommand{\AxK}{\mathtt{AxK}}
\newcommand{\AxT}{\mathtt{AxT}}

\newcommand{\AxTc}{\mathtt{AxT^{c}}}
\newcommand{\AxF}{\mathtt{Ax4}}

\newcommand{\AxFc}{\mathtt{Ax4^{c}}}
\newcommand{\AxV}{\mathtt{Ax5}}

\newcommand{\IndNatM}{\tspc
  \mathtt{Ind^{\raisebox{1pt}{\footnotesize{$\tspc \mathtt{m}$}}}_{\tspc \NN}} \tspc}
\newcommand{\IndNatL}{\tspc \mathtt{Ind^{\tspc \NN}_{\tspc\mathnormal{l}}} \tspc}
\newcommand{\modIndNat}{\tspc \mathtt{Ind^{\tspc \NN}_{\tspc\mathnormal{m}}} \tspc}
\newcommand{\IndNat}{\tspc \mathtt{Ind_{\tspc \NN}} \tspc}
\newcommand{\IndBool}{\tspc \mathtt{Ind_{\tspc \BB}} \tspc}

\newcommand{\realir}[1]{#1_{\BS \oplus}}
\newcommand{\refir}[1]{#1_{\BS \ominus}}
\newcommand{\contract}{\uplus}
\newcommand{\uzt}{ \BS \zeta }
\newcommand{\uztp}{ \BS { \prm{\zeta} } }
\newcommand{\uzts}{ \BS { \scd{\zeta} } }
\newcommand{\uztGama}{ \BS { \zeta^* } }
\newcommand{\Goedel}{\mbox{G\"odel}\xspace}
\newcommand{\Schuette}{\mbox{Sch\"utte}\xspace}
\newcommand{\Schuettes}{\mbox{Sch\"utte}'s\xspace}
\newcommand{\Eloise}{\mbox{Eloise}\xspace}
\newcommand{\Abelard}{\mbox{Abelard}\xspace}

\begin{document}
\begin{spacing}{1.1}

\title{Modal Functional (\texorpdfstring{``Dialectica''}{Dialectica}) Interpretation}

\author[D.~Hernest]{Dan Hernest}[a]
\address{Romanian Institute of Science and Technology, Cluj-Napoca, Romania}
\email{\url{danhernest@gmail.com}}
\thanks{The first author acknowledges support by the European Regional Development Fund and the Romanian Government through the Competitiveness Operational Programme 2014--2020, project ID P\_37\_679, MySMIS code 103319, contract no. 157/16.12.2016 .}

\author[T.~Trifonov]{Trifon Trifonov}[b]
\address{Faculty of Mathematics and Informatics, Sofia University ``St. Kliment Ohridski'', Sofia, Bulgaria}
\email{\url{triffon@fmi.uni-sofia.bg}}

\begin{abstract}
  We adapt our light Dialectica interpretation %
  to usual and light modal formulas (with universal quantification on boolean and natural variables) and prove it sound for a non-standard modal arithmetic based on \Goedel's $T$ and classical $S_4$. The range of this {\em light modal Dialectica} is the usual (non-modal) classical Arithmetic in all finite types (with booleans); the propositional kernel of its domain is Boolean and not $S_4$. The `heavy' modal Dialectica interpretation is a new technique, as it cannot be simulated within our previous light Dialectica. The synthesized functionals are at least as good as before, while the translation process is improved. Through our modal Dialectica, the existence of a realizer for the defining axiom of classical $S_5$ reduces to the Drinking Principle (cf. Smullyan).
\end{abstract}

\maketitle

Functional interpretations derived from \Goedel's computability adaptation \cite{Goedel(58)} of Aristotle's insights have been continuously developed over the years for constructive purposes. Modelizations and unified presentations abound \cite{DINIS2021102940}, as well as practical mathematical results from Kohlenbach's {Proof Mining} \cite{KohlenBook} continuation of Kreisel's {Unwinding of Proofs}.
When it comes to employing such proof interpretations for the synthesis of concrete computer code (certified by construction), only the quasi-direct reading of programs from already constructive proofs of input-output specifications has enjoyed a good measure of social success in academia (\eg\cite{Let2008}), while the industrial applications rather fall into the proof-carrying code paradigm (\eg\cite{marche04jlap}). Yet a good number of prototype examples have been worked out under the general umbrella of {\it program extraction from classical proofs} (\eg\cite{RAFFALLI200449, RatiuTrifonov}).

In \cite{dialfine}, the second author thoroughly presented how \Goedel's Dialectica interpretation can be completely deconstructed from its full computational essence down to a symbolic null transformation\footnote{\mssp See also Chapter 5 of \cite{TriffonPhD} for a more comprehensive exposition, in particular Section~5.5.1, page~129\mssp.}. However, the {\em flag} apparatus for
decorating\footnote{\mssp Note that in \cite{TriffonPhD} \RgRes characterizes full {\em lack of} computational content and corresponds to \NCRes here, \PlRes stands for partial content from the negative side and corresponds to \MnRes here, and \MnRes from \cite{TriffonPhD} denotes partial content from the positive side hence corresponds to \PlRes here. Basically polarities were reversed by the second author (already since \cite{dialfine}) due to his reconstructive approach which is otherwise dual (for quantifiers) to our constructive approach here. See also Footnote~2 on page~6 of \pCite{dialfine}
\label{FNTpolaritySwitch}} %
both quantifiers and implications (throughout the input proofs) tends to become too complex for human operators (so that Oliva's detour to the linear logic substructure \cite{Oliva[HFILIL]} may seem
a better alternative).

Here we propose a middle path between removing computational content of (`computationally correct') proofs via the second author's ``deep annotation'' mechanism and Oliva's ``shallow annotation'' equivalent approach (cf. Section~6 of \cite{dialfine}).
We will thus use \bmc{\necsy} a single switch, directly at the level of natural proofs. Although $\necsy$ cannot be simulated within our previous light Dialectica (hence is a strict addition to our previous light Arithmetic), it certainly is implementable within either of Trifonov's or Oliva's systems.

The purpose of our approach has been the rapid implementation in the actual \Minlog system (cf.~\cite{MinLogDoc} and Chapter~7 of \cmCite{pcbook} in particular Section~7.4). Indeed, $\necsy$ was implemented (cf.~\cite{MLFD}) as ``syntactic sugar'' over the `non-computational' implication \impnc seen as Kreisel implication.

Our modal systems are {\it normal} according to the definition from \cmCite{FittingMPF} and non-standard since the {\em normality scheme} $\AxK$ is (syntactically) derivable from the axiom scheme $\AxT$\mssp.

\section{Introduction}

The present work supersedes the functional synthesis technique outlined in our previous paper \cite{HerTrifonLDR} by adding a useful device for (homogeneously) combining the effect of previous optimizations by partly and fully uniform quantifiers in a compact releaser of constructive potential, namely the modal operator $\necsy$ (and its weak co-modality $\wpossy \Equiv \cpossy$). Proofs which are not necessarily {\it prima facie} constructive may yet potentially contain constructive content; in order to make use of this constructive `charge' contained in a (non-constructive) proof, various `release' instruments have been created over the past decades.

We will prove that $\necsy$ is not ``syntactic sugar'' over the functional interpretation of \cite{HerTrifonLDR}, but a genuinely new device (albeit synthesized out of previous works), cf. Section~\refp{SecBoxNoSS} We also bring the following result (cf. Theorem~\ref{ThmTUnrealS5}): {\it while the modal propositional axioms of system $S_4$ are realizable, the defining axiom of $S_5$ is not realizable, in general, under the modal functional interpretation, by primitive recursive functionals of finite type}.

The use and interpretation of modal operators in this paper were inspired by the work
of Oliva (partly joint with the first author, see \cite{HernestOliva}) at the linear
logic level, see \cite{Oliva(2007),Oliva[HFILIL]}.
It is no coincidence that, at formula level, our interpretation of $\necsy A$
is syntactically the same as Oliva's modified realizability interpretation of
$\lnecsy A$ in intuitionistic linear logic. However, a certain
detour would be needed in order to simulate $\necsy A$ in terms of $\lnecsy A$,
which may be less suitable for the processing of natural
proofs by humans (see Remark~1.23 in \cite{Girard(87B)}).

The second author independently noticed the possibility of using the same
supra-linear modal operators for light program extraction in \cite{dialfine},
see also \cite{TriffonPhD}. However, the initiative of studying the full employment
of $\necsy$ for more efficient functional synthesis in the formal context of the
negative fragment of first-order modal logic (cf. \Schuette \cite{Schuette} and
Prawitz \cite{Prawitz(65)}) is due to the first author.
As we will see, for our extractive purposes it is useful to depart from \Schuette's
original semantics for quantified modal logic. For example, the propositional
fragment of our first-order modal systems is not modal, but purely boolean,
as $\ \necsy p \,\EquiVl\, p \,\EquiVl \wpossy p\ $ for propositional atoms $p$.

We thus design two non-standard modal arithmetics, $\MSys \sset \tspc \MSysL$,
for functional program synthesis.
The soundness of these input systems is syntactically given via our (light) modal
functional interpretation by the target system, namely classical
decidable-predicate Arithmetic with higher-type functionals, in a Natural Deduction
presentation.\footnote{\mssp Note that soundness of \Schuette's predicate modal logics
(\mssp \eg \lbmh{S^{\star}_4}) is proved non-constructively,
using models, see \cite{Schuette} (\mssp cf. Chapter~I\mssp, $\S4$\mssp).}

For an easier exposition we will give up the \SQT{non-standard} prefix. Throughout
the paper, our modal Arithmetics are non-standard (relative to the conservative extensions
of $S_4$ due to Prawitz and \Schuette) but they resulted in a natural manner relative
to the Dialectica interpretation. It turns out that $\MSys$ intrinsically relates to the
{\em modally closed} subset of Prawitz's $\CSpFive$ (cf. \cite{Prawitz(65)}, page 77);
see also Remark~\refp{Rmkmvspsm}

Note that there was some attention to formalizing Quantified Modal Logic
stemming from Artificial Intelligence (cf. \cite{RQLKBMA}) and  there is a dedicated
Chapter~12 in \pCite{NegriPlatoPA}

\section{Arithmetical systems for light \AndSlashOr modal Dialectica extraction}\label{SeCbase}

We build upon functional arithmetical systems $\VSys$ and (the light annotated)
$\ISys$ from \pCite{HerTrifonLDR} While the \emph{verifying system} $\VSys$
basically is the Arithmetic $Z$ of Berger, Buchholz and Schwichtenberg
\cite{Berger(02)} in a slightly different presentation which is more suitable for light
functional synthesis and features classical logic (without strong existence) and full
extensionality\footnote{\mssp As inherited from system $Z$, our $\VSys$ is mostly
a Natural Deduction presentation of the so-called \SQT{negative arithmetic} from
\cite{Troelstra(73)}, basically a \Goedel-Gentzen embedding %
of classical into Heyting Arithmetic $\HAom$.},
its light counterpart $\ISys$ is only partly classical.

Moreover, the \emph{input system} $\ISys$ is weakly extensional and its contraction
(and hence also induction) rule is restricted for soundness of the (light)
functional interpretation of $\ISys$ into $\VSys$. In computing terms, the program
synthesis algorithm provided by the light Dialectica (of \cite{HerTrifonLDR}, as
inherited from the one\footnote{\mssp The restriction on extensionality is at its turn inherited
from \Goedel's functional interpretation (cf. \cite{Avigad(98)}, see also \cite{Goedel(58)}),
whereas the restriction on contraction was initially added by the first author in \cite{Hernest[PhD]},
as it was imposed by the necessity of decidability of the translation of light contraction formulas.}
of \cite{Hernest[PhD]}) produces correct output only modulo the
above-mentioned restrictions on Extensionality and Contraction\footnote{\mssp These
restrictions are more relaxed than those from the first author's PhD thesis
and weaker than \Goedel's restriction on extensionality, Kreisel's avoidance of
contraction in his Modified Realizability \cite{Kreisel(59)} and Girard's total
elimination of contraction in his original Linear Logic \cite{Girard(87B)}.}.
If not for the weak extensionality, $\ISys$ were a conservative extension of $\VSys$.

For (light) modal functional synthesis we will use the same verifying system $\VSys$.
The simpler input system $\MSys$ is obtained by adding $\necsy$ to a restricted variant
of $\VSys$. This (weakly extensional) modal Arithmetic will be proved sound via the
{\em modal Dialectica interpretation}. The fully-fledged input system $\MSysL$ adds
to $\MSys$ all light universal quantifiers and is a modal extension of $\ISys$;
its soundness will be given by the {\em light modal}  Dialectica interpretation.
Together with our new systems $\MSys$ and $\MSysL$ we will also present the relevant details
of arithmetics $\VSys$ and $\ISys$. Nonetheless for the full picture\footnote{\mssp %
In this paper we give a more detailed treatment of induction for numbers and
we correct the typo in the definition of %
$\RlCompat$:
on page 1382 of \cite{HerTrifonLDR}, it is $s$ instead of $x$ and $t$ instead of $y$\mssp,
cf. \ptref{DefCMPrule} and Section~\ref{SeCExtens}\mssp.}
we refer the reader to \cite{HerTrifonLDR} (see also \cite{dialfine} for a more complete picture).

We will use the same kind of \lND (\QT{\mssp ND\mssp}) presentation\mssp\footnote{\mssp
  A similar presentation style was employed by de Paiva in her categorical approach to linear logic (with modalities,
  see Sections~1.5 and 4.6 of \cite{VCVdePaivaPhD}), as imported from \cite{GirardLafontLLLC}\mssp.}
of our systems, where proofs are represented as sequents
\bmc{\spGamma \prfg B \tspc} meaning that formula $B$ is the root of the \ND tree
whose leaves $\spGamma$ are typed assumption variables (\QT{avars})
$\tspc \avar{a}{A} \tspc$. Here formula $\tspc A \tspc$ is the type of the avar
$\tspc a \tspc$ and $\spGamma$ is a multiset (since there may be more leaves
labeled with the same $\avar{a}{A}$, cf. \cite{Prawitz(65)}-Appendix C-$\S2$,
\QT{Variants of Gentzen-type systems}).

\begin{spacing}{1.2}
The sets of finite types \bmc{\Typ} terms \bmh{\Term} (of \Goedel's $\GT$)\mssp,
formulas $\Fmla$ (of $\VSys$) and $\FmlaL$ (of $\ISys$)\mssp, and\mssp,
with the addition of $\necsy$\mssp, formulas $\FmlaM$ of $\tspc \MSys \tspc$ and
$\FmlaML$ of $\tspc \MSysL \tspc$ are defined as follows:
\Array{@{\hspace{-4pt}}r@{\hspace{4pt}}c@{\hspace{1pt}}l@{\hspace{-35pt}}r}{
\Typ  & \dtRho, \dtSgm & \bndef \NN \bnor \BB \bnor (\dtRho \dtSgm) & \\[10pt]
\Term  & s, t & \multicolumn{2}{@{\hspace{1pt}}l}{\bndef x^{\dtRho} \bnor \BolTru^{\BB} \bnor \BlFals^{\BB}
\bnor \Zero^{\NN} \bnor \Succ^{\NN \NN} \bnor \If^{\BB \dtRho \dtRho \dtRho}
\bnor \Rec^{\NN \dtRho (\NN \dtRho \dtRho) \dtRho}
\bnor (\lbda{x^{\dtRho}}{ t^{\dtSgm}})^{\dtRho \dtSgm}
\bnor (t^{\dtRho \dtSgm} s^{\dtRho})^{\dtSgm}} \\[10pt]
\Fmla  & A, B & \bndef \atom{t^{\BB}}
\bnor A \dtimp B \bnor A \dtand B \bnor \fa x^{\dtRho} A
& \framebox[1.05\width]{$\LogFalsity \, \EqBDL \,
\atom{\BlFals}$, $\dtnot A \; \EqBDL \; A \dtimp \LogFalsity$} \\[10pt]
\FmlaL & A,B &\bndef \atom{t^{\BB}}
\bnor A \dtimp B \bnor A \dtand B \bnor \fa x^{\dtRho} A
\bnor \, \fa_{\set{\yset,+,-}} x^{\dtRho} A
& \framebox[1.05\width]{\,$\ex x^{\dtRho} A \, \EqBDL \,
\dtnot \fa x^{\dtRho} \dtnot A$\,} \\[10pt]
\FmlaM  & A, B & \bndef \atom{t^{\BB}}
\bnor A \dtimp B \bnor A \dtand B \bnor \fa x^{\dtRho} A \bnor \necsy A
& \framebox[1.05\width]{\,$\wpossy A \; \EqBDL \; \cpossy A$\,} \\[10pt]
\FmlaML & A,B &\multicolumn{2}{@{\hspace{1pt}}l}{\bndef \atom{t^{\BB}}
\bnor A \dtimp B \bnor A \dtand B \bnor \fa x^{\dtRho} A
\bnor \necsy A \bnor \, \fa_{\set{\yset \tspc , \tspc + \tspc , \tspc -}} x^{\dtRho} A}\\[8pt]
}
For simplicity we employ two basic types: integers $\NN$ and booleans
\bmc{\BB} and use $\tspc \dtRho \tspc \dtSgm \tspc \dtTau \tspc$ for
$\tspc ( \tspc \dtRho \tspc ( \tspc \dtSgm \tspc \dtTau \tspc) \tspc ) \tspc$.
Building blocks for terms are the constructors for booleans
$\lbrack \BolTru, \BlFals \rbrack$ (\mssp{\em true} and {\em false}, both of type
$\BB$), integers $\lbrack \Zero, \Succ \rbrack$ (\mssp{\em zero}, of type $\NN$ and
{\em successor}, of type $\NN \NN$), \Tpymphc case distinction
$\tspc \If \tspc$ and \Tpymphc \Goedel recursion $\tspc \Rec \tspc$.

Atomic formulas $\atom{t^{\BB}} $ are decidable by definition, as they are
identified with boolean terms $t^{\BB}$. In particular, we have decidable falsity
\bmh{\LogFalsity \nspc \EqBD \nspc \atom{\BlFals}}
and truth \bmp{\LogTruth \nspc \EqBD \nspc \atom{\BolTru}}
We abbreviate \bmh{A \dtimp \LogFalsity} by \bmp{\dtnot A}
The partially light universal quantifiers $\fapl$, $\famn$ (partly computational)
and $\fanc$ (non-computational) are inherited from \cite{HerTrifonLDR}.

The universal quantifier $\fa$, axiomatized as usual in Natural Deduction,
will have full computational content in the input systems. The
weak existential quantifier $\tspc \ex \tspc$ is defined for formulas in all
our systems as $\, \ex x^{\dtRho} A \; \EqBDL \; \dtnot \fa x^{\dtRho} \dtnot A \tspc$.
The weak co-modality operator $\tspc \wpossy \tspc$ is defined
for formulas in $\tspc \FmlaM \tspc$ and $\; \FmlaML \;$
as $\ \wpossy A \; \EqBDL \; \cpossy A \tspc$.
\end{spacing}

We purposefully avoid specifying types for terms insofar they can be deduced from the meta-context. In all our systems, the meta-operator $\FV{\cdot}$ will return the set of free variables of its argument, which can be a term or a formula.

\subsection*{Term system \texorpdfstring{$\Term$}{T}}

Computation in our systems is expressed by means of the usual $\beta$-reduction
rule \bmc{(\lambda x.t)s \converts t [x\mapsto s]} together with the rewrite
rules defining the computational meaning of $\If$ and \bmcc{\Rec}
\Array{rlrl}{
 \If\,\BolTru\,s\,t\;&\converts\;s&
\qquad\qquad\Rec\,\Zero\,s\,t\;&\converts\;s\\[6pt]
 \If\,\BlFals\,s\,t\;&\converts\;t&\qquad\qquad
\Rec\,(\Succ n)\,s\,t\;&\converts\;t\,n\,(\Rec\,n\,s\,t)
}
Since this typed term system is confluent and strongly normalizing
(cf. Section~6.2.5 of \cite{pcbook}), we are free not to fix a particular evaluation
strategy.

For simplicity, we will assume
that all terms occurring in our formal proofs automatically get into normal form,
as normalization is necessary only when matching terms in formulas. We thus avoid
introducing equality axioms like in \cite{Hernest[PhD]} and skip the corresponding
easy applications of extensionality.
In conclusion, some computations get to be carried out implicitly when building proofs in our systems\mssp\footnote{\mssp This is just \Minlog's mechanism, cf. \cite{MinLogDoc}, see also \cite{MLFD} for our personalized distribution\mssp.}\mssp.

Using recursion at higher types we can define any provably total function of
ground arithmetic, including decidable predicates such as
equality $\EqBool$ for booleans and $\EqNatt$ for natural numbers:
\Array{r@{\quad}c@{\quad}l}{
\EqBool^{\BB \BB \BB}&\EqBD&
\lbda{x}{\If \, x \, ( \lbda{y}{y} ) \,
( \lbda{y}{\If \, y \, \BlFals \, \BolTru} )}\\[8pt]
\EqNatt^{\NN \NN \BB} &\EqBD& \lbda{x}{ \Rec \, x\, \big( \lbda{y}{
\Rec \, y \, \BolTru \, ( \lbda{n, q^{\BB}}{\BlFals})} \big) \, \big( \lbda{
m, p^{\NN \BB}, y}{ \Rec \, y \, \BlFals \, ( \lbda{n, q^{\BB}}{p \, n} )} \big) }
}
\begin{table}[!p]
\begin{center}
  \begin{tabular}{|@{\quad}c@{\qqquad}c@{\quad}|}\hline&\\[-8pt]
    \mArray{r@{:\quad }l}{
      \AxCompat & \prfg\  x=_{\dtRho} y \, \dtimp \, A(x) \, \dtimp \, A(y)\\[10pt]
      \AxTrue & \prfg\ \atom{\BolTru} }&
    \pt \Gmnsama\ \lprg\ s=_{\dtRho} t \jst \Gmnsama\ \lprg\ B(s)
    \dtimp B(t) \usg \RlCompat_{\dtRho} \ept \\[14pt]
\hline\end{tabular}
\end{center}
\caption{Basic axioms, with $\AxCompat$ replaced by
$\RlCompat$ rule in $\ISys$, see \ptref{DefCMPrule} and Section~\ref{SeCExtens}}
\label{Axioms}
\end{table}

\begin{table}[!p]\vspace{18pt}
\begin{center}
\begin{tabular}{|@{\;\;}c@{\quad}c@{\quad}c@{\quad}|@{\quad}c@{\;\;}|}
\hline&&&\\[-6pt]
$ \avar{a}{A} \prfg A \quad \id$&
\pt \spGamma, [\avar{a}{A}] \prfg B \jst \spGamma \prfg A \dtimp B \usg \LimpI \ept&
\pt \spGamma \prfg A \quad \spDelta \prfg A \dtimp B \jst
\spGamma, \spDelta \prfg B \usg \LimpE \ept&
\pt \spGamma \prfg  A \jst \spGamma \prfg \Bfa z \nspc A \usg \faIB \ept\\[24pt]
\pt \spGamma \prfg A \dtand B \jst \spGamma \prfg A\usg \landEl \ept&
\pt \spDelta \prfg A \land B \jst \spDelta \prfg B \usg \landEr \ept&
\pt \spGamma \prfg A \quad \spDelta \prfg B \jst
\spGamma, \spDelta \prfg A \land B \usg \landI \ept&
\pt \spGamma \prfg \Bfa z \nspc A \jst \spGamma \prfg  A[z\mapsto t] \usg \faEB \ept\\[14pt]
\hline\end{tabular}
\end{center}
\caption{Logical rules, with $\tspc z \NotIn \FV{\spGamma} \tspc$ at $\faIB$ and
  contractions due to $\tspc \LimpE \tspc$ and $\tspc \landI \tspc$ explicitated as
  anti-rules, see Table~\ref{Con-rules}; no implicit contractions at $\LimpI$}
\label{VSys-rules}
\end{table}

\begin{table}[!p]\vspace{18pt}
\begin{center}
\begin{tabular}{|@{\quad}c@{\quad}c@{\quad}|}\hline&\\[-6pt]
\begin{tabular}{c@{\quad\ and\ \quad}c}
\pt \spGamma \lprg  A \jst \spGamma \lprg \faplh z \; A \usg \faplhIB \ept&
\pt \spGamma \lprg \faplh z \; A \jst \spGamma \lprg  A[z\mapsto t] \usg \faplhEB \ept%
\end{tabular}
&for\quad $\plh\in \set{\yset \tspc , \tspc + \tspc , \tspc -}$\\[14pt]
\hline\end{tabular}
\end{center}
\caption{Additional rules for \bmc{\ISys} with extra restrictions
on \bmc{\faplIB} \bmh{\famnIB} and \bmc{\fancIB}\,\
see\,\ \cmPlRes \MnRes and \NCRes in Section~\ref{SeCISys}}
\label{ISys-rules}
\end{table}

\begin{table}[!p]\vspace{18pt}
\begin{center}
\begin{tabular}{|@{\quad}c@{\qqquad}c@{\quad}|}\hline&\\[-6pt]
\pt \spDelta, \avar{a}{A}, \avar{a}{A} \prfg B
\jst \spDelta, \avar{a}{A} \prfg B \usg \, \mCnRl \ept&
\pt \spDelta, \avar{a}{A}, \avar{a}{A} \lprg B
\jst \spDelta, \avar{a}{A} \lprg B \usg \, \mLCnRl \ept\\[14pt]
\hline\end{tabular}
\end{center}
 \caption{Contraction anti-rules \mssp\CnRlTx\mssp for $\VSys$ and
($\bigstar$\mssp-\mssp restricted) \mssp\LCnRlTx\mssp for $\ISys$, see Remark~\ref{RmkContract}}
\label{Con-rules}
\end{table}

\begin{table}[!p]\vspace{18pt}
\begin{center}
\begin{tabular}{|@{\quad}c@{\qquad}c@{\quad}|}\hline&\\[-6pt]
\pt \spGamma \prfg A(\BolTru) \quad \spDelta \prfg A(\BlFals)
\jst \spGamma, \spDelta \prfg A(b) \usg \IndBool \ept&
\pt \spGamma \prfg A(\Zero) \quad \spDelta \prfg A(n) \dtimp A(\Succ
n) \jst \spGamma, \spDelta \prfg A(n) \usg \IndNat \ept\\[24pt]
\multicolumn{2}{|@{\qquad}c@{\qquad}|}{\pt \spGamma\ \lprg\ A(\Zero) \qquad \spDelta \dspc \lprg \dspc
A(n) \; \dtimp \; A (\Succ n) \jst \spGamma \contract \spDelta \dspc
\lprg \dspc A(n)\usg{\ \IndNatL}\ept}\\[14pt]
\hline\end{tabular}
\end{center}
 \caption{Induction rules, with $\tspc \spGamma \contract \spDelta \tspc$
instead of \mssp\SQT{$\spGamma, \spDelta$}\mssp and $\tspc \spDelta \tspc$
restricted via $\bigstar$ at the induction over numbers of $\tspc \ISys$, \ie \bmc{\IndNatL}
see Section~\ref{SeCIndNat}}
\label{Ind-rules}
\end{table}
\subsection{The verifying system \texorpdfstring{$\VSys$}{NA}}\label{SeCVSys}

The logical rules of system $\VSys$ are presented in Table~\refc{VSys-rules}
with the usual restriction on $\faIB$ (universal quantifier introduction)  that
\Array{r@{\quad\,}c@{\quad\,}l}{
z & \NotIn &\FV{\spGamma} \; \EqBDL \
\BS{\bigcup}_{\textstyle \tspc \avar{a}{A}  \in \spGamma}\, \  \FV{A}
}

At \bmc{\LimpI} $[\avar{a}{A}]$ denotes the unique occurrence
of $\avar{a}{A}$ in the multiset of assumptions of the premise sequent of
\bmp{\LimpI} Thus \bmc{\avar{a}{A}\NotIn\spGamma} hence $\avar{a}{A}$ is
no longer an assumption in the conclusion sequent of \bmp{\LimpI} In the usual
tree representation of Natural Deduction proofs, the leaf
labeled \QT{$\avar{a}{A}$} gets inactivated\mssp\footnote{\mssp
  Or \QT{discharged}, as one usually says in Natural Deduction terminology\mssp.}\mssp,
after (possibly) multiple of its copies had (all) been equalized to it via
instances of the {\it contraction anti-rule} (henceforth called ``contractions'').

While for $\VSys$ itself one could allow that all contractions be handled
implicitly at \bmc{\LimpI} in relationship with the architecture of light input systems
(e.g.,\mssp $\ISys$, cf. Section~\ref{SeCISys}) we are compelled to introduce for $\VSys$
the contraction anti-rule \CnRlTx in association with the corresponding
\LCnRlTx (of, e.g.,\mssp $\ISys$, cf. Table~\ref{Con-rules})\mssp.

We refer to contraction as \QT{anti-rule}, rather than \QT{rule} because,
despite the sequent-like representation of our calculi, in fact our formalisms
are \ND and in the \ND directed tree the representation of explicit contractions
is by convergent arrows that go in the direction which is reverse to the
direction of all the other rules\footnote{\mssp Sequentwise though, contraction is a rule,
cf. pages 90\mssp,\mssp91 of \cite{Prawitz(65)}-A-$\S1$,$\S2$\mssp.}\mssp.

We find it convenient to introduce induction for booleans and numbers as the rules
presented in Table~\refp{Ind-rules} Here we assume that the induction variables
$\,b^{\BB}\,$ and respectively $\,n^{\NN}\,$ do not occur freely in \bmc{\,\spGamma}
nor \bmc{\,\spDelta} and that they do occur in the formula \bmp{\tspc A}

The $\atom\cdot$ construction allows us to view boolean programs as decidable
predicates. Given \bmc{\IndBool} its logical meaning is settled by the truth axiom
\bmc{\AxTrue} see Table~\refp{Axioms} In this way we can define predicate equality
at base types as \Array{rcl}{ s \tspc =_{\dtSgm} \tspc t \ &\EqBD&\
\atom{ \Eq_{\!\dtSgm} \, s \; t }\quad \mbox{ \ for \  $\dtSgm \in \set{\BB,\NN}$}}
and further at higher types, extensionally, as %
\Array{rcl}{ s \tspc =_{\dtRho\dtTau} \tspc t \ & \EqBD &
\ \fa x^{\dtRho} \, ( \tspc s \tspc x \, =_{\dtTau} \, t \tspc x \tspc) }
It is straightforward to prove by induction on
$ \, \dtRho \, $ that $ \, \ =_{\dtRho} \; $ is reflexive,
symmetric and transitive at any type \bmp{\tspc \dtRho}

To complete our system, we include in \lbmh{\VSys} also the compatibility
(i.e., extensionality) axiom \bmc{\AxCompat} see Table~\refp{Axioms}
Note that ex falso quodlibet (\lbmh{\EFQ}) \lbmh{\tspc \LogFalsity \dtimp A \nspc}
and stability (\lbmh{\Stab}) \lbmh{\tspc \twonot A \tspc \dtimp \tspc A \nspc} are
fully provable in \lbmh{\VSys} (cf. Section~1.4 of \cmCite{TriffonPhD} by induction on the
logical structure of \bmc{A} using \lbmh{\AxTrue} and \lbmh{\IndBool}, see also Chapter~1 of \cite{pcbook}
or \cite{MinLogDoc}--10.6)\mssp.

\subsection{Input system \texorpdfstring{$\ISys$}{NAl}}\label{SeCISys}%
Light formulas $\FmlaL$ were built over usual formulas $\Fmla$ of $\VSys$ by adding three\footnote{\mssp For the universal quantification with combined positive/negative computational content we here use $\fa$ instead of the more verbose $\forall_{\!\pm}$ from \cmCite{HerTrifonLDR} as it should be clear from the meta-context whether an actual instance of $\fa$ is in an input proof (hence part of $\ISys$) or a verifying proof (thus part of $\VSys$).} light universal quantifiers: the non-computational $\fanc$ and the two semi-computational $\fapl$ and $\famn$ (see also Footnote~\ref{FNTpolaritySwitch})\mssp.

Thus, system $\ISys$ refined the adaptation of $\VSys$ (with $\RlCompat$ for $\AxCompat$ and \LCnRlTx for \CnRlTx) with introduction and elimination rules for the light quantifiers (see Table~\ref{ISys-rules}). These are copies of the regular \ND rules $\faEB$ and \bmc{\faIB} but with the usual restriction on \llbmh{\faIB}  that \Lbmh{z\NotIn\FV{\spGamma}} enhanced with the following
conditions\mssp\footnote{\mssp Restrictions \PlRes, \MnRes and \NCRes assume in-depth knowledge of subproofs, so that
input proofs are defined inductively in parallel with the extraction of part of
their computational content (namely free variables of already synthesized terms).}
referring to the interpretation of \llbmh{\spGamma \lprg  A}:

\begin{itemize}[left=4mm]\label{Extra_restrictions}
\item[\PlRes]
in the \lbmh{\faplIB\tspc} rule, \lbmh{z} may be used computationally only positively, \ie \lbmh{z} must not be free in the \emph{challengers} of the translation of \llbmh{\spGamma} (\mssp basically \lbmh{z \NotIn \cup_{i=1}^n \tspc \FV{\!\ut_i\!}}, cf. Statement~\ref{soundness}\mssp)
\item[\MnRes]
in the \lbmh{\famnIB\tspc} rule, \lbmh{z} may be used computationally only negatively, \ie \lbmh{z} must not be free in the \emph{witnesses} of the translation of \llbmh{A} (\mssp cf. Example~\ref{DfLDT}\mssp; basically \lbmh{z \NotIn \FV{\!\ut_0\!}}\mssp)
\item[\NCRes]
in the \lbmh{\fancIB\tspc} rule, \lbmh{z} may not be used computationally at all,
\ie  both \,\PlRes\,\ and \,\MnRes\mssp.
\end{itemize}

Classes of \emph{realization irrelevant} $\realir{A}$ and \emph{refutation irrelevant} $\refir{A}$ formulas\footnote{\mssp A formula is realization irrelevant iff its tuple of witness variables is empty. A formula is refutation irrelevant iff its tuple of challenge variables is empty. See the equivalent Remark~1 in Section~3 of \cite{HerTrifonLDR}.\label{fntR1S3}} are defined as follows (below $\nothing$ denotes no thing):
\Array{rcl}{
\realir A,\realir B &\bndef& \atom{t}\bnor \realir A \dtand
\realir B \bnor \refir A \dtimp \realir B \bnor \FaPlh x \;
\realir A\quad \mbox{ for \ } \plh \in \set{\yset,+,-,\nothing}\\[8pt]
\refir A,\refir B &\bndef& \atom{t} \bnor \refir A \dtand
\refir B \bnor \realir A \dtimp \refir B \bnor \FaPlh x \; \refir A
\quad \mbox{ for \ } \plh \in \set{\yset,+}
}

Since Dialectica is unable to interpret full extensionality (cf. \cite{KohlenbSpect,Troelstra(73)})
one has to replace $\AxCompat$ with a weak compatibility rule.
We thus employ an upgraded variant of the \Tpymphc $\RlCompat$ rule from
\cite{Hernest[PhD]} (herewith called {\it light extensionality}):
\begin{equation}\label{DefCMPrule}
\pt \Gmnsama\ \lprg\ s=_{\dtRho} t \jst \Gmnsama\ \lprg\ B(s)
\dtimp B(t) \usg \RlCompat_{\dtRho} \ept
\end{equation}
where all formulas in $\Gmnsama$ are refutation irrelevant, \ie the negative (challenge)
position in their translation (cf. Section \ref{SeCLDI} below) is empty.

The computationally irrelevant contractions of $\ISys$ (i.e., whose formula is refutation irrelevant)
can\footnote{\mssp This was an instrumental %
  compromise between the first author's implementation with tuples (cf. \cite{Hernest[PhD]})
  and the second author's implementation with pairs (cf. \cite{MinLogDoc, TriffonPhD}, see also Section~7.4 of \cite{pcbook}).} %
be handled implicitly at \bmp{\LimpI} The situation is different for those contractions
whose formula is refutation relevant (i.e., the {\em computationally relevant contractions}),
as we wanted to automatically ensure that their translation is decidable (instead
of leaving the task of decidability check to the user, as we shall for the upcoming modal systems).

The decidability of their translation is necessary for attaining soundness.

\begin{rem}[restriction $\bigstar$ on relevant contractions]\label{RmkContract}
  We achieve a decidable translation by including in $\ISys$ the contraction anti-rule \LCnRlTx
  (see Table~\ref{Con-rules})
  where $\bigstar$\mssp: all formulas $A$ that are refutation relevant {\it must not contain
  any \bmc{\fapl} nor \bmh{\fanc}}\mssp. This triggered the addition to \bmh{\VSys} of an
  explicit (unrestricted) contraction anti-rule \CnRlTx which is needed in the construction
  of the verifying proof (it only applies to quantifier-free formulas $\ldIntB{A}$).
\end{rem}
We thus ensured that all contraction formulas that require at least one challenger term for their
light interpretation would have quantifier-free
(hence decidable) translations\mssp\footnote{\mssp %
For the (light) modal Dialectica we will upgrade this purely syntactical
criterion used in \cite{HerTrifonLDR} (as inherited from \cite{Hernest[PhD]}),
see Definition~\ref{DefContR} at the end of Section~\ref{ModSys}\mssp.}\mssp. %
In \cite{HerTrifonLDR}, in order to avoid having to deal with any computationally relevant
contractions implicitly at \bmc{\LimpI} we had constrained the deduction rules of $\ISys$ to
disallow multiple occurrences of refutation relevant assumptions
in any of the premise sequents\mssp\footnote{\mssp Thus, whenever a double occurrence of a
refutation relevant assumption were created in a conclusion sequent by one
of the binary rules of \bmc{\ISys} such sequent could not be directly a premise
for the application of an(other) $\ISys$ rule: the anti-rule  \LCnRlTx
had to be applied first, in order to eliminate the critical double\mssp.}\mssp.

We here no longer need such an explicit constraint, given the stronger (yet equivalent) implicit
constraint imposed by the requirement at $\LimpI$ that the cancelled assumption $\avar{a}{A}$
is a singleton. It is thus left to the implementation to lean towards lazy handling of
contractions (all gathered just before $\LimpI$, suitable for parallel execution within
eager environments, as hinted by \cite{Hernest[PhD]}) or the second author's \cite{TriffonPhD}
eager handling of contractions (so that assumptions basically form a set) that turned out
to be better suited for the lazy evaluation paradigm, or anything in-between\footnote{\mssp
  A {\em monotone} variant (cf. \cite{kohlenbach_1992}, see also \cite{KohlenBook}) would
  not care much of where to handle relevant contractions, as it benefits from their easy
  realization via simple (default, or at most user provided) majorants\mssp.}\mssp.

While \bmh{ \nspc \EFQ: \, \LogFalsity \tspc \dtimp \tspc A \nspc} remains fully
provable also in $\ISys$ (for all formulas $A\in\FmlaL$) the situation changes
for \bmh{ \nspc \Stab: \, \twonot A \tspc \dtimp \tspc A \nspc}
in the case of many formulas $A$ that feature light quantifiers in certain places\footnote{\mssp
  As outlined in Section~3.1 of \cite{HerTrifonLDR} and noted already in \cite{Hernest[PhD]},
  the usual proof in $\VSys$ of $\Stab$ (constructed by induction on $A$) unavoidably makes
  use of contractions over \bmh{\neg\neg (B \dtand C)} for subformulas \bmh{(B \dtand C)}
  of \bmc{A} and these are subject to the $\bigstar$ restriction for refutation relevant
  \bmp{B \dtand C} Even when such subformulas %
  do obey $\bigstar$, they may lead to the failure of restrictions \PlRes, \MnRes or \NCRes.}\mssp.

On the other hand, $\Stab$ is provable in $\ISys$ for $A\in\Fmla$ or $A$ conjunction-free.

\subsection{Light functional interpretations}\label{SeCLDI}

Any formula $A$ of an input system is translated to a not necessarily
quantifier-free formula \bmh{ \InLdInt{A}{\ux}{\uy} } of $\VSys$ so that
$\ux, \uy$ are tuples of fresh (not appearing in \bmh{A}) variables.
The $\ux$ in the superscript are the \emph{witness variables},
while subscript variables $\uy$ are the \emph{challenge variables}.

Terms $\ut$ substituting witness variables (\mssp like
\bmh{\InLdInt{A}{\ut}{\uy}}) are called \emph{realizing terms}
or \QT{witnesses} and terms $\us$ substituting challenge variables
(\mssp like \bmh{\InLdInt{A}{\ux}{\us}}) are called \emph{refuting terms}
or \QT{challengers}.
The interpretation of specification $A$ can be seen as a game\mssp\footnote{\mssp We acquired the
  game semantics interpretation (originating in  \cite{BLASS1992183}) from
  works of Oliva\mssp.} in which {\Eloise} ($\iex$) first and then {\Abelard}
($\fa$) make one move each by  playing objects $\ut$ and $\us$ of corresponding types
for the tuples $\ux$ and respectively \bmp{\uy}

Formula $\InLdInt{A}{\ux}{\uy}$ specifies the not necessarily decidable
(as it were for \Goedel's Dialectica) \QT{adjudication relation}. Eloise wins iff
\bmp{\, \VSys \prfg \InLdInt{A}{\ut}{\us} \tspc}

\begin{exa}[Definition of light Dialectica translation of formulas, from \cite{HerTrifonLDR}]\,\\[0pt]
\label{DfLDT}\indent
The interpretation preserves atomic formulas, \ie
\bmp{ \ldIntB{ \atom{ t^{\BB} } } \EqBDL \atom{ t^{\BB} } }
\mssp Assuming \bmh{ \, \InLdInt{A}{\ux}{\uy} \, }
and \bmh{ \, \InLdInt{B}{\uu}{\uv} \, } are already defined,
\Array{r@{\quad}c@{\quad}l@{\;\;\mbox{and}\;\;}r@{\quad}c@{\quad}l}{
\ldInt{A \dtand B}{\ux, \uu}{\uy,\uv} & \EqBD &
\ldInt{A}{\ux}{\uy} \dtand \ldInt{B}{\uu}{\uv}&
\ldInt{A \dtimp B}{ \uf, \ug }{ \ux, \uv } & \EqBD &
\ldInt{A}{\ux}{\uf \ux \uv} \dtimp
\ldInt{B}{\ug \ux}{\uv}
}
The interpretation of the four universal quantifiers is (upon renaming,
we assume that quantified variables occur uniquely in a formula):
\Array{l@{\quad}c@{\quad}l@{\qqquad}l@{\quad}c@{\quad}l}{
\ldInt{\fa z \; A(z)}{\uh}{z,\uy} & \EqBD &
\ldInt{A(z)}{\uh z}{\uy}&
\ldInt{\fapl z \; A(z)}{\uh}{\uy} & \EqBD &
\fa z \nspc \ldInt{A(z)}{\uh z}{\uy}\\[10pt]
\ldInt{\famn z \; A(z)}{\ux}{z,\uy} & \EqBD &
\ldInt{A(z)}{\ux}{\uy}&
\ldInt{\fanc z \; A(z)}{\ux}{\uy} & \EqBD &
\fa z \nspc \ldInt{A(z)}{\ux}{\uy}
}
Since $\ldIntB{\LogFalsity} \EquivL \LogFalsity$ we get
\mArray{l@{\,}c@{\,}l}{\ldInt{\dtnot B}{\uV}{\uu} & \Equiv & \dtnot \ldInt{B}{\uu}{\uV \uu}}
hence \mArray{l@{\,}c@{\,}l}{\ldInt{\twonot A}{\uX}{\uY} & \Equiv &
\twonot \ldInt{A}{\uX \uY}{\uY (\uX \uY)}} and also
\Array{l@{\;}c@{\;}l@{\;\quad}l@{\;}c@{\;}l}{
\ldInt{\dtnot \tspc \fa z \;\tspc A(z)}{\tspc Z\tspc,\uY}{\uh} & \Equiv &
\dtnot \ldInt{A(\tspc Z \uh)}{\uh \tspc (\tspc Z\tspc \uh)}{\uY \uh}&%
\ldInt{\dtnot \tspc \famn z \;\tspc A(z)}{\tspc Z\tspc,\uY}{\ux} & \Equiv &
\dtnot \ldInt{A(Z \ux)}{\ux}{\uY \ux}\\[10pt]
\ldInt{\dtnot \tspc \fapl z \;\tspc A(z)}{\uY}{\uh} & \Equiv &
\dtnot \fa z \, \ldInt{A(z)}{\uh \tspc z}{\uY \uh}&%
\ldInt{\dtnot \tspc \fanc z \;\tspc A(z)}{\uY}{\ux} & \Equiv &
\dtnot \fa z \, \ldInt{A(z)}{\ux}{\uY \ux}
}
It is straightforward to compute (\mssp for weak existential counterparts
\lbmh{\ExPlh \tspc x \EqBDl \lnot \tspc \FaPlh x \tspc \lnot \tspc} with
\lbmh{\tspc \plh \in \set{\eset\dtcsp+\dtcsp-\dtcsp }}\mssp) that
\Array{@{\hspace{0pt}}l@{\!}c@{\,}l@{\ }l@{\!}c@{\,}l}{
\ldInt{\exf z \; A(z)}{\ZtCmUh}{\uYs} & \Equiv &
\twonot \ldInt{\AZuYs}{\uH \uYs}{ \uYs \tspc ( \tspc Z \uYs ) \nspc ( \uH \uYs ) }&
\ldInt{\expl z \; A(z)}{\uH}{\uYs} & \Equiv &
\ex z \; \ldInt{A(z)}{\uH \uYs}{ \uYs \tspc z \nspc ( \uH \uYs ) }\\[12pt]
\ldInt{\exmn z \; A(z)}{\ZtCmUh}{\uYs} & \Equiv &
\twonot \ldInt{\AZuYs}{\uH \uYs}{ \uYs ( \uH \uYs ) }&
\ldInt{\exnc z \; A(z)}{\uH}{\uYs} & \Equiv &
\ex z \; \ldInt{A(z)}{\uH \uYs}{ \uYs ( \uH \uYs ) }
}%
The length and types of the witnessing and challenging tuples are uniquely determined for a given
formula\mssp. \dtLBrack Note that cf. Definition~\refc{DfMDI}
\mArray{rcl}{\ldInt{\necsy\fa z \; A(z)}{\uh}{}&
  \equiv&\fa z,\uy\,\ldInt{A(z)}{\uh z}{\uy}} \dtRBrack%
\end{exa}

{\Eloise} will have a winning move whenever specification $A$ is provable in the input system:
the light interpretation will explicitly provide it from the proof of \bmc{A} as a tuple of witnesses
\bmh{\ut} \mbox{\lbrack\ such that  $\tspc \FV{\ut} \; \sseq \ \FV{A}$ \rbrack}
together with the \emph{verifying proof} in \bmh{\VSys} of
\bmh{\nspc \fa \! \uy \tspc \InLdInt{A}{\ut}{\uy} \tspc}
(\mssp Eloise wins by $\ut$ regardless of the instances
$\us$ for {\Abelard}'s $\uy$).

The following parameterized statement gives a practical pattern in which soundness theorems for Dialectica-based interpretations can uniformly be expressed in a \ND setting. The metavariables $\metaISys$ and $\metaVSys$ below stand for input and respectively verifying systems.

\begin{sta}[generic soundness for Dialectica interpretations
\mtparam{\metaISys}{\metaVSys}]\label{soundness}
Let \lbmh{A_0 \dtcsp A_1 \dtcsp \ldots \dtcsp A_n} be a sequence of formulas
of \lbmh{\metaISys} with $\uw$ all their free variables. If the sequent
\Lbmh{\avar{a_1}{A_1} \dtcsp \ldots \dtcsp \avar{a_n}{A_n}\;\lprg\; A_0}
is provable in \lbmc{\metaISys} then terms
\lbmh{\ut_0 \dtcsp \ldots \tspc, \ut_n} can be automatically synthesized
from its formal proof, such that the translated sequent
\Array{r@{\quad}c@{\quad}l}{
\avar{a_1}{\ldInt{A_1}{\ux_1}{\ut_1}}, \ldots,
\avar{a_n}{\ldInt{A_n}{\ux_n}{\ut_n}}&\prfg&\ldInt{A_0}{\ut_0}{\ux_0}
}
is provable in \bmc{\metaVSys} and the following \emph{free variable condition}
\FVC holds: \bmh{\ux_0\NotIn\FV{\ut_0}} and \bmp{\FV{\ut_i} \sseq\{\uw, \ux_0, \ldots, \ux_n\}} Here \bmh{\ux_0, \ldots, \ux_n} are tuples of fresh variables, such that equal avars share a common such tuple.
\end{sta}

In \cite{HerTrifonLDR} the above was thoroughly proved for $\metaISys\equiv\ISys$ and $\metaVSys\equiv\VSys$, except for the interpretation of $\RlCompat$ %
which we present below. Further in the sequel we also give a %
more detailed treatment of the induction rule for numbers, in order to motivate the introduction of the {\em modal induction rule} in Section~\refp{SecModIndRule}

\subsection{Light Extensionality}\label{SeCExtens}

We here give the interpretation of \ptref{DefCMPrule}. By definition of equality at higher types,
\bmh{\, s \tspc =_{\dtRho}  r \;} is
\bmc{\; \fa \! \uz \tspc . \ s \uz \tspc = \tspc r \uz}
hence a purely universal formula. We are given that
\Array{rcl}{\avar{a_1}{\ldInt{A_1}{\ux_1}{\ut_1}}, \ldots,
\avar{a_n}{\ldInt{A_n}{\ux_n}{\ut_n}} & \prfg & \ldInt{A_0}{\ut_0}{\ux_0}\quad ,\\[6pt]}
where $\, \ldIntB{\Gmnsama} \tspc \Equiv \tspc \set{a_1, \ldots , a_n} \tspc$,
$\tspc \ut_0 \tspc \Equiv \tspc \ut_1 \tspc \Equiv \tspc
\ldots \tspc \ut_n \tspc \Equiv \tspc \etup \tspc$
(empty tuple)\mssp, $\tspc A_0 \tspc$ is $\tspc s \tspc =_{\dtRho} \tspc r \tspc $
and $\ux_0$ corresponds to $\uz$, thus the above is more conveniently rewritten as
\Array{rcl}{\avar{a_1}{\ldInt{A_1}{\ux_1}{}}, \nspc \ldots \nspc ,
\avar{a_n}{\ldInt{A_n}{\ux_n}{}} & \prfg & s \, \ux_0 = r \, \ux_0 \,}
To this we can apply the generalization rule, as $\ux_0$ are not free in the
translated context $\ldIntB{\Gmnsama}$. Indeed, $\ux_0$ are fresh variables
and they could have appeared free only via terms
$\tspc \ut_1 \tspc ,\tspc \ldots \tspc , \tspc \ut_n \tspc$, were these not
empty tuples (hence the need for restricting the original context). We thus obtain
$\tspc \ldIntB{\Gmnsama} \, \prfg \, s = r \tspc$ and further apply $\AxCompat$ to get
$\tspc \ldIntB{\Gmnsama} \, \prfg \, \ldIntB{B}(s) \dtimp \ldIntB{B}(r) \tspc$.
Note that the \ul{axiom} is required here, as $\ldIntB{\Gmnsama}$ may contain general\mssp\footnote{\mssp
  The verification in a $\metaVSys$ with Spector's \ul{rule} of extensionality (instead of axiom),
  employed as $\RlCompat$ in our framework, would already fail for $\Pi^0_1$ assumptions in $\Gmnsama$,
  as first discovered by Kohlenbach in \cite{KohlenbSpect}.
}  formulas.

With $\tspc g \EqBDL \lbda{\uu}{\uu} \tspc$ and $\nspc f \EqBDL \lbda{\uu, \uv}{\uv} \tspc$
we have thus constructed a verifying proof
\Array{@{\hspace{-1pt}}r@{\;}c@{\;}l}{
\avar{a_1}{\ldInt{A_1}{\ux_1}{}}, \nspc \ldots \nspc ,
\avar{a_n}{\ldInt{A_n}{\ux_n}{}} & \prfg &
\ldInt{B(s)}{\uu}{f \uu \uv} \dtimp \ldInt{B(r)}{g \uu}{\uv}
\; \big\lbrack \, \Equiv\ \,
\ldInt{B(s) \dtimp B(r)}{f \dtcsp g}{\uu, \uv}
\, \big\rbrack
\,}
The new realizing terms \lbmh{f \dtcsp g} are closed, hence the free variable condition
trivially holds.

Note that \lbmh{f} and \lbmh{g} may at most depend on the type
\lbmh{\rho} {\em (\mssp they do not depend on concrete terms \lbmh{s \dtcsp r})}\mssp,
see also the first example in Section~\refp{SeCEx}

\subsection{Numbers}\label{SeCIndNat}
Since the induction rule (for numbers, see Table~\ref{Ind-rules}) corresponds to an unbounded number
of contractions of each assumption from the step context $\spDelta$ (cf. \cite{Hernest[PhD]}), its
clone in the system $\ISys$ is subject to a restriction like the one of \LCnRlTx.
Namely, we need to require that {\em all refutation relevant avars in $\spDelta$ satisfy}
$\tspc\bigstar$ (cf. Remark~\ref{RmkContract}).

Moreover, since the contractions on $a \in \spGamma \cap \spDelta$ will be handled
differently than for simple binary rules like $\LimpE$ or \bmc{\landI} it is more
convenient to require that induction over numbers in $\ISys$ implicitly contracts all its
refutation relevant assumptions (instead of using the explicit \LCnRlTx).
We will use the notation $\spGamma \contract \spDelta$ for a special multiset union
in which refutation relevant assumptions appear only once, even if they appear in
both $\spGamma$ and \bmp{\spDelta}

Thus the $\IndNatL$ rule of $\ISys$ is finally obtained by replacing \SQT{$\spGamma, \spDelta$}
with \SQT{$\spGamma \contract \spDelta$} in the conclusion sequent of \bmp{\IndNat} %
For the verifying proof, we are given
\begin{eqnarray}\label{EqIndBase}
\ldInt{\spGamma}{\uu}{\ugm[\uy]}&\prfg&\ldInt{A(\Zero)}{\ur}{\uy}\\[4pt]
\label{EqIndStep}
\ldInt{\spDelta}{\uz}{\udt[\ux;\uv]}&\prfg&
\ldInt{A(n)}{\ux}{\ut\ux\uv}\ \dtimp\ \ldInt{A(\Succ n)}{\us\ux}{\uv}
\end{eqnarray}
We show that
\begin{equation}\label{AuxInd}
\fa\uv\nspc\big(\,\ldInt{\spGamma\contract\spDelta}{\,
\uu\contract\uz}{\,\uzt[n]\tspc\uv}\;\dtimp\;\ldInt{A(n)}{\utp[n]}{\uv}\,\big)
\end{equation}
is a theorem of \Lbmc{\VSys} where
\begin{equation}\label{EqIndTP}
\tp[n]\ \EqBDL\ \Rec \; n \; r \; ( \lbda{n}{s} )
\end{equation}
for every corresponding pair \lbmh{\pair{r \in \ur}{s \in \us}} and \lbmh{\uzt[n]}
will be constructed as functional terms  depending on \lbmp{\uv}
We here intentionally use the same variable \lbmh{n} that occurs freely in \lbmh{s}
and \lbmp{t} Implicitly, just \lbmh{\tp} denotes \lbmp{\tp \tspc [n]} Also \lbmh{\uzt}
will be constructed as the collection of all \lbmh{\uztp} (corresponding to
\lbmh{\spGamma \setminus \spDelta}) and \lbmh{\uzts}
(corresponding to \lbmh{\spDelta}).
Here $\uu\contract\uz$ denotes the tuple union corresponding to the multiset union $\spGamma\contract\spDelta$, i.e.,
witness variables corresponding to refutation relevant assumptions in $\spGamma\cap\spDelta$ appear only once.

Let \lbmh{\avar{b}{B}} be a refutation relevant avar in
\lbmp{\spGamma \contract \spDelta}\,
Let \llbmh{\ugmp \in \ugm} \AndSlashOr \llbmh{\udtp \in \udt} be the challengers for \llbmh{b}
in \llbmh{\spGamma} \AndSlashOr \lbmp{\spDelta} If \llbmh{b} appears only in \lbmh{\spGamma}
(hence not in \lbmh{\spDelta}) we define
\begin{eqnarray}\label{EqIndZTP}
\uztp[n] & \EqBD & \Rec \; n \; (\tspc \lbda{\uv}{\ugmp[\uv]} \tspc) \;
\big(\tspc \lbda{\tspc n \tspc , \tspc p, \uv}{p\,(\tspc \uttpv \tspc)} \tspc \big)
\end{eqnarray}

If $\tspc b\tspc$ appears in \bmc{\spDelta} then the decidability of \bmh{\ldIntB{B}}
is needed at each recursive step to equalize the terms \Lbmh{p \nspc (\uttpv)}
obtained by the recursive call with the corresponding terms \bmp{\udtp} Thus the
right stop point of the backwards construction is provided. In fact an implicit
contraction over \lbmh{b} happens at each inductive step and $\bigstar$ guarantees
that \bmh{\ldIntB{B}} is decidable.

For \Lbmh{b\in\spGamma\cap\spDelta} let
\begin{equation}\label{EqIndOne}
  \uzts[n] \ \EqBD \
  \Rec \nspc n \nspc \big( \lbda{\uv}{ \ugmp [ \uv ]} \big)
\nspc \Big( \, \lbda{ \tspc n \tspc , \tspc p \tspc , \uv}{ \tspc
\If \nspc ( \nspc \ldInt{B}{\uzp}{ \udtp [ \utp ; \uv ] } \tspc )
\nspc  \big( \tspc p \tspc ( \uttpv ) \tspc \big) \nspc  \udtp[ \utp ; \uv ]
} \; \Big)
\end{equation}
and for \Lbmh{ b \in \spDelta \setminus \spGamma } we define its
\Lbmh{\uzts[\tspc n \tspc ] \tspc}
by replacing in \reffl{EqIndOne} the \lbmh{\tspc \ugmp} with canonical zeros.
Here \lbmh{\uzp} are the challenge variables corresponding to formula
\lbmp{B} Notice that
\begin{eqnarray}\label{EqIndTwo}
& \prfg &
\nspc \tp \nspc [ \nspc \Succ \tspc n \nspc ]
\dspc = \dspc
s \  \tp \tspc [ \tspc n \tspc ]\\[3pt]\label{EqIndThree}
& \prfg &
\nspc \uztp \nspc [ \nspc \Succ \tspc n \nspc ] \nspc \uv
\dspc = \dspc
\uztp [ \tspc n \tspc ] \; ( \uttpv )\\[3pt]\label{EqIndNew}
&\prfg&
\nspc \uzts \nspc [ \nspc \Succ \tspc n \nspc ] \nspc \uv
\dspc = \dspc
\If \,\  ( \tspc \ldInt{B}{\uzp}{ \udtp [ \utp ; \uv ] } \tspc ) \,\
\big( \, \uzts \tspc [ \tspc n \tspc ] \dspc ( \uttpv ) \, \big)
\,\  \udtp [ \utp ; \uv ]
\end{eqnarray}
We attempt to extend \reffL{EqIndThree} to the whole \lbmh{\uzt}
by proving from \reffL{EqIndNew} the following
\begin{eqnarray}\label{EqIndFour}
\ldInt{B}{\uzp}{ \uztsSnUV } \ & \prfg & \
\uztsSnUV \dspc = \dspc \uztsNuttpv
\end{eqnarray}
We obtain this as an immediate consequence of
\begin{eqnarray}\label{EqIndFive}
\ldInt{B}{ \uzp }{ \uztsSnUV }
 & \prfg & \ldInt{B}{ \uzp }{ \UdtpUtpUv }
\end{eqnarray}
Assuming \Lbmc{\, \neg \ldInt{B}{ \uzp }{ \UdtpUtpUv } \tspc}
\ by \reffL{EqIndNew} we get\Array{rcl}{
\, \uztsSnUV \ = \ \UdtpUtpUv \,,&\mbox{hence}&
\, \neg \ldInt{B}{ \uzp }{ \uztsSnUV} \,}
and thus \reffL{EqIndFive} follows via \llbmh{\Stab}
(which is fully available in the verifying system).

We now prove \reffL{AuxInd} by an assumptionless
induction on \lbmp{n} Let \lbmh{\uztGama} be the collection
of all \lbmh{\uztp} and those \lbmh{\uzts} corresponding to
\lbmp{ \spGamma \cap \spDelta }\,\
For \llbmh{ n \EquiVl \Zero } it is sufficient that
\begin{eqnarray*}
\ldInt{\spGamma}{\uu}{ \uztGamZ \tspc \uv } & \prfg &
\ldInt{ A ( \Zero ) }{ \utpZero }{ \uv }
\end{eqnarray*}
which follows  from \reffL{EqIndBase} since by definition \reffL{EqIndTP}
we have \llbmh{ \prfg \, \utpZero \tspc = \tspc \ur} and by
definitions \reffL{EqIndZTP} and \reffL{EqIndOne} we have
\Lbmp{ \prfg \, \uztGamZ  \tspc = \tspc \lbda{\uv}{ \ugm [ \uv ] } }
\,\ Now given \reffL{AuxInd} we want to prove
\begin{eqnarray}\label{AuxIndBis}
\ldInt{\spGamma\contract\spDelta}{\,
\uu \contract \uz }{ \, \uztSucN \tspc \uv }
& \prfg & \ldInt{ A( \Succ n ) }{ \utpSucN }{ \uv }
\end{eqnarray}
To \reffL{AuxInd} we apply \Lbmh{ \faE{ \uv \, \mapsto \, \uttpv } } and
via easy deductions in \lbmh{\VSys} we get
\begin{eqnarray}\label{EqIndSix}
\ldInt{\spGamma\contract\spDelta}{ \, \uu \contract \uz }{
\, \uztN \nspc ( \uttpv ) } & \prfg & \ldInt{A(n)}{ \utpN }{ \uttpv }
\end{eqnarray}
With \reffL{EqIndThree} and \reffL{EqIndFour} we can rewrite
\reffL{EqIndSix} to
\begin{eqnarray}\label{EqIndSeven}
\ldInt{\spGamma\contract\spDelta}{\,\uu\contract\uz}{\,\uztSnUV}
& \prfg & \ldInt{A(n)}{\utpN}{\uttpv}
\end{eqnarray}
In \reffL{EqIndStep} we substitute
\Lbmh{\ux\,\mapsto\,\utp[n]}  and get
\begin{eqnarray*}
\ldInt{\spDelta}{\uz}{\udt[\utp;\uv]}&\prfg&\ldInt{A(n)}{\utp[n]}{
\ut\utp\uv}\,\dtimp\,\ldInt{A(\Succ n)}{\us\utp[n]}{\uv}
\end{eqnarray*}
which  gives \reffL{AuxIndBis} by means of easy \lbmh{\VSys} deductions
using \reffLc{EqIndTwo} \reffL{EqIndFive} and \reffLp{EqIndSeven}

\subsection{Motivation for the modal induction rule}\label{sec_motiv}

We have treated the most general situation, with all context sets
\lbmc{\spGamma\setminus\spDelta} \lbmh{\spGamma\cap\spDelta} and
\lbmh{\spDelta\setminus\spGamma} inhabited by refutation relevant assumptions,
and conclusion formula \lbmh{A} accepting both witnesses and challengers.

Many particular situations amount to easier treatments, with simpler
extracted terms. These can be obtained as simplifications of the general
witnesses and challengers presented above, by means of the reduction
properties of the empty tuple \bmh{\varepsilon} (practically the same as
for the isomorphic \emph{nullterm} from Section~7.2.4 of \cmCite{pcbook} also denoted $\varepsilon$).

We outline below only those particular cases which are relevant in connection
with the modal induction rule \bmh{\IndNatM} (cf. Section~\ref{SecModIndRule})\mssp:
\begin{itemize}
\item
If \llbmh{\spGamma\cup\spDelta} contains no refutation relevant assumption,
but \lbmh{A(n)} is refutation relevant, then terms \lbmh{\ut} are not part
of the realizers  for the conclusion sequent, in this case only $\utp$.
Hence \lbmh{\ut} would be redundantly produced and a mechanism is
needed to prevent their construction. This is ensured by \lbmh{\necsy} in
front of the step \lbmh{A(n)} at \lbmp{\IndNatM}%
\item
If \Lbmh{A(n)} is refutation relevant, \lbmh{\spDelta} has no refutation
relevant element but \lbmh{\spGamma} is refutation relevant inhabited, then
\lbmh{\udt} and \lbmh{\uzts} are empty. Yet \lbmh{\uztGama \EquivL \tspc \uztp}
has to be produced as \ptref{EqIndZTP} and includes \bmsc{\ut[n]} \mssp this
no longer will be the case for \lbmh{\IndNatM} (cf. technical details at the end
of Section~\ref{SecModIndRule} further in the sequel; challengers $\ugm$ simply
are preserved for $\ldIntB{\spGamma}$)\mssp.%
\item
If \Lbmh{A(n)} is refutation irrelevant then \lbmc{\uv} \lbmh{\ut} and
\lbmh{\uttpv} are empty tuples. Thus \lbmh{\uztp \EquiVl \ugmp} and
\reffL{EqIndOne} simplifies to\\[1pt] [ recall that \lbmc{ n \NotIn \FV{\ugmp} }
\lbmc{ n \in \FV{\utp} } and possibly \bmh{\, n \in \FV{\udtp} } ]
\begin{eqnarray*}%
\uzts [ \tspc n \tspc ] & \EquiVl & \Rec \; n \; \ugmp \, \Big( \,
\lbda{\tspc n \tspc , \tspc p \tspc}{ \If \, \big( \ldInt{B}{\uzp}{\udtp [ \utp ]} \big)
\ p \ \udtp [ \utp ] } \, \Big)
\end{eqnarray*}
\end{itemize}

\section{Modal system \texorpdfstring{$\MSys$}{NAm} and light modal system \texorpdfstring{$\MSysL$}{NAml}}\label{ModSys}

The usual propositional restriction on the introduction rule for the necessity operator is
that all contextual assumptions had been discharged prior to the rule application
(which amounts to forcing $\tspc \spGamma \tEquiv \yset \tspc$ at standard \InecsYsp).
In the natural deduction presentation of standard modal logic, \InecsY cannot be
unrestricted or $\tspc A \dtimp \necsy A \tspc$ becomes a theorem, thus all occurrences
of $\tspc\necsy\tspc$ becoming redundant.

Our restriction on \InecsY is strictly weaker, as, e.g., allows any context $\spGamma$
whose formulas are all refutation irrelevant
(this is akin to Prawitz's `first version' in \cite{Prawitz(65)}VI.$\S1$)
and any context at all if the conclusion is  refutation irrelevant.
Thus, $\tspc A \dtimp \necsy A\tspc $ not only is more generally possible in our
quantified modal systems, it even defines a quite interesting class of formulas,
see Definition~\ref{DfNecsyFmla}\mssp.

We %
polymorphically use the `proof gate' $\mprfg$ for both $\MSys$ and $\MSysL$, and use $\mlprg$
to stress that the proof belongs to \bmp{\MSysL} The constraints outlined below the
tables on page~\lpgref{VSys-rules} smoothly adapt to the
insertion of $\necsy$ (into the input system $\ISys$, through $\mINcsy$ and $\AxT$), eventually followed by
the removal of $\famn$, $\fapl$ and $\fanc$, and also to the upgrade from $\bigstar$ to $\maltese$\mssp,
as described in the sequel (cf. new tables on page \lpgref{mod_VSys-rules}, with $\modCnRl$ for $\mLCnRl$ and $\modIndNat$ for $\IndNatL$).

For the necessity operator $\necsy$ we have the following {\em enhanced} introduction rule, which admits many more premise sequents than usual (\mssp as the context $\spGamma$ may be inhabited\mssp)\mssp:
\[ \mINcsy:\ \quad \pt \spGamma \mprfg A \jst \DB{\spGamma \mprfg \necsy A} \ept \quad,\]
where $\spGamma$ is restricted depending on the (light) modal translation of the proof of \lbmh{A} from $\spGamma$, in a way that is akin to the condition \mssp\PlRes \mssp on the \lbmh{\faplIB} rule from page \lpgref{SeCISys}; see Definition~\ref{DfNecIntro} further below.

The following axioms of modal propositional logic $S_4$ (\mssp cf. \cite{Schuette}, Chapter~VII; see also Chapter~9 of \cite{Troelstra(96)}\mssp) are part of $\MSys$ and $\MSysL$:
\Array{l@{\qquad\qquad\qquad}l}{
\AxT:\ \necsy A \dtimp A & \AxTc:\  A \dtimp \wpossy A\\[8pt]
\AxF:\ \necsy A \dtimp \necsy \necsy A &
\AxFc: \wpossy \wpossy  A \dtimp \wpossy  A\\[8pt]
\multicolumn{2}{l}{\AxK:\ \lbrack \,
\necsy(A \dtimp B) \dtand \necsy A \, \rbrack \dtimp \necsy B}
}
\noindent In fact only \bmh{\AxT} is needed as an axiom of our non-standard modal systems.
Of course, \lbmh{\AxTc} and \lbmh{\AxFc} had been syntactically deducible from
\lbmh{\AxT} and respectively \lbmh{\AxF}  already in the propositional modal
system \lbmc{S_4} only using
minimal logic (\mssp the proof of \bmh{\AxFc} also uses \bmh{\AxK} and the empty-context
\InecsYb\mssp). It turns out that also $\AxF$ and $\AxK$ are easily deducible
in $\MSys$/$\nspc\MSysL$ just from $\AxT$ (and only using minimal logic),
given our very liberal necessity introduction rule, see Definition~\ref{DfNecIntro}
below.

Note that Stability $\twonot B \,\dtimp\, B$ needs to be restricted
already for $\MSys$, due to the necessary restriction on Contraction, cf. Definition
\lref{DefContR} in the sequel, see also Remark~\refp{Rmkmvspsm}

We denote by \Lbmh{A \kimp B\ \EqBD\ \necsy A \dtimp B} the so called \SQT{Kreisel implication}\footnote{\mssp See Section~3.2 of \cite{Oliva[HFILIL]} for a sketch of this construct and its design difficulties within the multi-modal linear setting. See also  \cite{Prawitz(65)}, Chapter~VII \QT{some other concepts of implication} for a discussion on notions of stronger implication which appeared  since early research on modal logic.},
since its translation  by (light) modal Dialectica is akin to its Modified Realizability
interpretation. Basically, if \lbmh{A} is a formula in which all implications are Kreisel ones, then the modal Dialectica  interpretation of \lbmh{\necsy A} is logically equivalent (provably in $\VSys$) to the modified realizability interpretation of \mssp$A$\mssp; \mssp see Lemma 3.2 of \cite{Oliva(06)} and also \cite{Oliva(2012)[UFI-PF]}\mssp.

Note that even though our Kreisel implication looks similar to the so-called `lax implication' (cf. \cite{PfenningDavies}, Section~7), here we are not concerned with a standard (intuitionistic) modal logic (see Remark~\lref{Rmkmvspsm} at the end of Section \lref{SeCLMD}). Ditto for the (classical) translation of \lbmh{\necsy} under the Curry-Howard-style modal functional interpretation of De Queiroz and Gabbay (cf. \cite{deQueirozGabbay1997}, see also Section 7 of \cite{de2012functional} for an updated survey)\mssp.

\begin{table}[!p]
\begin{center}
  \begin{tabular}{|@{\quad}c@{\qqquad}c@{\quad}|}\hline&\\[-6pt]
    \mArray{r@{:\quad }l}{
    \AxT&\mprfg\ \necsy A \dtimp A\\[8pt]
    \AxTrue&\mprfg\ \LogTruth}&
    \pt \Gmnsama\ \mprfg\ s=_{\dtRho} t \jst
    \DB{\Gmnsama\ \mprfg\ B(s) \dtimp B(t)} \usg \RlCompat_{\dtRho}\ept\\[16pt]
  \hline\end{tabular}
\end{center}
\caption{Axioms of $\MSys$ and $\MSysL$, and light extensionality \ptref{DefCMPrule} adapted cf. Remark~\ref{rmk_LMT}}
\label{mod_Axioms}
\end{table}

\begin{table}[!p]\vspace{18pt}
\begin{center}
\begin{tabular}{|@{\:\:\,}c@{\:\:\,}c@{\:\:\,}c@{\:\:\,}|@{\:\:\,}c@{\:\:\,}|}
\hline&&&\\[-6pt]
$ \avar{a}{A} \mprfg A \quad \id$&
\pt \spGamma, [\avar{a}{A}] \mprfg B \jst \DB{\spGamma \mprfg A \dtimp B} \usg \LimpI \ept&
\pt \spGamma \mprfg A \quad \spDelta \mprfg A \dtimp B \jst
\DB{\spGamma, \spDelta \mprfg B} \usg \LimpE \ept&
\pt \spGamma \mprfg  A \jst \DB{\spGamma \mprfg \Bfa z \nspc A} \usg \faIB \ept\\[24pt]
\pt \spGamma \mprfg A \dtand B \jst \DB{\spGamma \mprfg A} \usg \landEl \ept&
\pt \spDelta \mprfg A \land B \jst \DB{\spDelta \mprfg B} \usg \landEr \ept&
\pt \spGamma \mprfg A \quad \spDelta \mprfg B \jst
\DB{\spGamma, \spDelta \mprfg A \land B} \usg \landI \ept&
\pt \spGamma \mprfg \Bfa z \nspc A \jst \DB{\spGamma \mprfg  A[z\mapsto t]} \usg \faEB \ept\\[16pt]
\hline\end{tabular}
\end{center}
\caption{Logical rules of $\MSys$ and $\MSysL$, with $\tspc z \NotIn \FV{\spGamma} \tspc$
  at $\faIB$ and contractions due to $\tspc \LimpE \tspc$ and $\tspc \landI \tspc$ explicitated
  as anti-rules, see Table~\ref{mod_Con-rules}; no implicit contractions at $\LimpI$}
\label{mod_VSys-rules}
\end{table}

\begin{table}[!p]\vspace{18pt}
\begin{center}
\begin{tabular}{|@{\quad}c@{\qquad}c@{\quad}|}\hline&\\[-6pt]
\begin{tabular}{c@{\quad\ and\ \quad}c}%
\pt \spGamma \mlprg  A \jst \DB{\spGamma \mlprg \faplh z \; A} \usg \faplhIB \ept &
\pt \spGamma \mlprg \faplh z \; A \jst \DB{\spGamma \mlprg  A[z\mapsto t]} \usg \faplhEB \ept\\[16pt]
\end{tabular}
&for\quad $\plh\in \set{\yset \tspc , \tspc + \tspc , \tspc -}$\\[8pt]
\hline\end{tabular}
\end{center}
\caption{Additional (relative to $\MSys$) rules for \bmh{\MSysL} with the (adapted, cf. Remark~\ref{rmk_LMT})
  extra restrictions on \bmc{\faplIB} \bmh{\famnIB} and \bmh{\fancIB}\,\ as in Section~\ref{SeCISys},
  cf.\,\ \cmPlRes \MnRes and \NCRes }
\label{mod_ISys-rules}
\end{table}

\begin{table}[!p]\vspace{18pt}
\begin{center}
\begin{tabular}{|@{\quad}c@{\qqquad}c@{\quad}|}\hline&\\[-6pt]
\pt \spGamma \mprfg A \jst \DB{\spGamma \mprfg \necsy A} \usg \, \mINcsy \ept&
\pt \spDelta, \avar{a}{A}, \avar{a}{A} \mprfg B
\jst \DB{\spDelta, \avar{a}{A} \mprfg B} \usg \, \modCnRl \ept\\[16pt]
\hline\end{tabular}
\end{center}
\caption{Necessity introduction rule with $\spGamma$ restricted via Definition~\ref{DfNecIntro}
  and contraction anti-rule $\modCnRl$ with $A$
  $\tspc\maltese\tspc$\mssp-\mssp restricted through
  Definition~\ref{DefContR}\mssp, for $\MSys$ and $\MSysL$}
\label{mod_Con-rules}
\end{table}

\begin{table}[!p]\vspace{18pt}
\begin{center}
\begin{tabular}{|@{\quad}c@{\quad\;}c@{\quad}|}\hline&\\[-6pt]
\pt \spGamma \mprfg A(\BolTru) \quad \spDelta \mprfg A(\BlFals)
\jst \DB{\spGamma, \spDelta \mprfg A(b)} \usg \IndBool \ept&
\pt \spGamma\ \mprfg\ A(\Zero) \qquad \spDelta \dspc \mprfg \dspc
A(n) \; \dtimp \; A (\Succ n) \jst \DB{\spGamma \contract \spDelta \dspc
\mprfg \dspc A(n)} \usg{\ \modIndNat}\ept\\[16pt]
\hline\end{tabular}
\end{center}
\caption{Induction rules of $\MSys$ and $\MSysL$, with $\spDelta$ of $\modIndNat$ restricted
  via the $\tspc\maltese\tspc$ upgrade (cf. Definition~\ref{DefContR}) of $\bigstar$
  (cf. Remark~\ref{RmkContract}), see Sections~\refc{SeCIndNat} \ref{sec_motiv} and \ref{SecModIndRule}}
\label{mod_Ind-rules}
\end{table}

\begin{defi}[modal  Dialectica interpretation --- translation of formulas]\ \\%
\label{DfMDI}%
The interpretation does not change atomic\mssp\footnote{\mssp Any decidable formula %
can (and should) be given via its associated boolean term, \eg one should rather use \bmh{\atom{\!\Odd{x}\!}} instead of the more verbose \bmc{\Bfa y \,(\tspc 2 \tspc y \tspc \neq \tspc x \tspc)} which is refutation relevant in a somewhat artificial and probably unintended way.\label{FntDF}} %
formulas, \ie \bmp{\ldIntB{\atom{t^{\BB}}} \, \EqBDL \, \atom{t^{\BB}}}\\[2pt]
Assuming \bmh{\ldInt{A}{\ux}{\uy}} and \bmh{\ldInt{B}{\uu}{\uv}} are already defined\mssp,
\Array{r@{\quad}c@{\quad}l@{\quad}r@{\quad}c@{\quad}l}{
\ldInt{A \dtand B}{\ux, \uu}{\uy , \uv} & \EqBD &
\ldInt{A}{\ux}{\uy} \dtand \ldInt{B}{\uu}{\uv}&
\ldInt{\Bfa z \nspc A(z)}{\uh}{z\tspc, \uy} & \EqBD &
\ldInt{A(z)}{\uh z}{\uy}\\[10pt]
\ldInt{A \dtimp B}{\uf, \ug}{\ux, \uv} & \EqBD &
\ldInt{A}{\ux}{\uf \ux \uv} \dtimp
\ldInt{B}{\ug \ux}{\uv}&
\ldInt{\necsy A}{\ux}{}& \EqBD &
\Bfa \uy \tspc \ldInt{A}{\ux}{\uy}
}\end{defi}\noindent
As an immediate consequence\mssp,
$$\inArray{r@{\ }c@{\ \,}l}{
\ldInt{\necsy \fa z \tspc A(z)}{\uh}{} & \EquivL &
\fa z \tspc,\uy \tspc \ldInt{A(z)}{\uh \tspc z}{\uy}
}\ ,\ %
\inArray{r@{\ }c@{\ \,}l}{
\ldInt{\lnot \necsy B}{}{\uu} & \EquivL & \lnot \fa \uv \ldInt{B}{\uu}{\uv}
}$$ and further %
\Array{r@{\quad}c@{\quad}l}{
\ldInt{\wpossy A \ \EquivL  (\tspc \cpossy A \tspc)}{}{\uf}
& \EquiVl &  \ex \ux \ldInt{A}{\ux}{\uf \ux} \\[12pt]
\ldInt{A \kimp B \ \EquivL (\tspc \necsy A \dtimp B \tspc)}{\ug}{\ux, \uv}
& \EquiVl &
\Bfa \uy \tspc \ldInt{A}{\ux}{\uy} \; \dtimp \; \ldInt{B}{\ug \ux}{\uv}\\[8pt]
\ldInt{\lnot \tspc \necsy \fa z \tspc A(z)}{}{\uh} & \EquivL &
\lnot \fa z \tspc,\uy \tspc \ldInt{A(z)}{\uh \tspc z}{\uy} } %
Recall from Example~\ref{DfLDT} in Section~\ref{SeCLDI} that
\dtLBrack recall that $\, \ex z \, A(z) \; \EqBDL \;
\dtnot \fa z \tspc \dtnot \tspc A(z) \tspc$ \dtRBrack
\Array{r@{\ }c@{\ \,}l}{%
\ldInt{\exf z \; A(z)}{\ZtCmUh}{\uY} & \EquivL &
\twonot \ldInt{\AZuY}{\uH \uY}{ \uY \tspc ( \tspc Z \uY ) \nspc ( \uH \uY ) }
}%
which we can compare with
\mArray{r@{\ }c@{\ \,}l@{\ }c@{\ \,}l}{%
\ldInt{\exf z \, \necsy \, A(z)}{z\tspc,\ux}{} & \EquivL & \twonot \ldInt{A(z)}{\ux}{}
                                          & \loquiv_{\VSys} & \ldInt{A(z)}{\ux}{}
}%
\\ or even
\Array{r@{\ }c@{\ \,}l}{%
  \ldInt{\necsy \exf z \; A(z)}{\ZtCmUh}{} & \EquivL & \fa \uY \twonot
  \ldInt{\AZuY}{\uH \uY}{ \uY \tspc ( \tspc Z \uY ) \nspc ( \uH \uY )}\\[6pt]&\loquiv_{\VSys}
& \fa \uY \tspc \ldInt{\AZuY}{\uH \uY}{ \uY \tspc ( \tspc Z \uY ) \nspc ( \uH \uY ) }
}

\begin{spacing}{1.2}
\begin{defi}[Necessity Introduction]\label{DfNecIntro}
The restriction on \mssp\InecsY\mssp is relative to programs synthesized from the proof
of the premise \llbmh{A} of this \mbox{Natural Deduction} rule, unless all formulas in the %
context \llbmh{\spGamma} are refutation irrelevant or \llbmh{A} is refutation irrelevant.
Namely, with \lbmh{\spGamma \Equiv \set{\avar{a_1}{A_1}, \ldots, \avar{a_n}{A_n}}}
and \lbmh{A \EquivL A_0}, the restriction is that
$\ux_0 \NotIn \cup_{i=1}^n \tspc \FV{\!\ut_i\!}$
in the translated premise sequent
\bmp{\avar{a_1}{\ldInt{A_1}{\ux_1}{\ut_1}}, \ldots,
  \avar{a_n}{\ldInt{A_n}{\ux_n}{\ut_n}}\prfg\ldInt{A_0}{\ut_0}{\ux_0}\,}
\end{defi}

Thus {\it admissible} input proofs are {\it inductively defined} together with their
extracted programs and their corresponding translated
(verifying) proofs.
Note that $\necsy$ could be defined in terms of $\kimp$ as
\lbmc{\necsy A \EquivL (A \kimp \bot) \dtimp \bot}
since $\VSys$ features full stability $\Stab$.
\begin{defi}[light modal Dialectica translation of formulas]
\label{DfLMDI}
The following are added to the above Definition~\ref{DfMDI} (\mssp the deduced translation of
\bmh{\exnc z} is outlined below for use at the end of Section~\ref{SeCEx}\mssp; see also
the proposed intuitionistic extension in Section~\ref{SeCFut}\mssp)\mssp:
\Array{r@{\ \ }c@{\ \ }l@{\qquad}r@{\ \ }c@{\ \ }l}{
\ldInt{\fapl z \  A(z)}{\uh}{\uy} & \EqBDl &
\fa z \ \ldInt{A(z)}{\uh z}{\uy}&
\ldInt{\famn z \  A(z)}{\ux}{z, \uy} & \EqBDl &
\ldInt{A(z)}{\ux}{\uy}\\[10pt]
\ldInt{\fanc z \  A(z)}{\ux}{\uy} & \EqBDl &
\fa z \ \ldInt{A(z)}{\ux}{\uy}&
\ldInt{\exnc z \  B(z)}{\uU}{\uV} & \EquiVl &
\ex z \ \ldInt{B(z)}{\uU \uV}{\uV (\uU \uV)}\\[-24pt]
}%
\end{defi}
\end{spacing}
\begin{rem}\label{rmk_LMT}
The light modal translation of formulas only adds
$\InLdInt{\necsy A}{\ux}{} \EqBDL \ \Bfa \! \uy \InLdInt{A}{\ux}{\uy}$
to our light translation from \cite{HerTrifonLDR}
(cf. Section~\ref{SeCbase} of this paper, in particular Example~\ref{DfLDT}).
\end{rem}

Formula \lbmh{A} is {\it realization relevant}
also under (light) modal Dialectica if the tuple of witness variables $\ux$
of its translation $\InLdInt{A}{\ux}{\uy}$ is not empty and similarly
\lbmh{A} is {\it refutation relevant} if the tuple of challenge variables
$\uy$ is not empty (see also Footnote~\ref{fntR1S3}\mssp)\mssp.

Correspondingly, \lbmh{A} is realization irrelevant if it is not
realization relevant (i.e., $\ux$ is an empty tuple), and \lbmh{A} is
refutation irrelevant if it is not refutation relevant
(i.e., $\uy$ is an empty tuple). \ \dtLBrack See also the more technical
definition in Section~\lref{SeCISys} \dtRBrack

\begin{rem}[restriction violation for \InecsY]\label{RmkRestrictViolBoxIntro}
In an automatized interactive search for modal input proofs of some given specification, we can temporarily allow unrestricted (or lesser restricted) instances of\ \InecsYsp\;and postpone the validity check for when the proof of its premise is fully constructed. This approach would be similar to the so-called \SQT{computationally correct proofs} mechanism of \cmCite{TriffonPhD} or \SQT{nc-violations} check since pre-\textit{decorate} \Minlog versions.
\end{rem}

For efficiency reasons, we recommend the use of modal operators whenever possible instead of the above partly (or non) computational quantifiers $\fapl$, $\famn$, $\fanc$ and $\exnc$. %
It thus makes sense to study the (pure) modal Dialectica in itself, as the use of such light quantifiers may not be needed in many
cases of interest. %

It should be easier to construct a strictly modal (i.e., without light quantifiers) input proof, also for a (semi) automated proof-search algorithm.
Nevertheless, it is the light variant of modal Dialectica which provides the larger range of possibilities, particularly for situations where the simpler, \SQT{heavier} modal Dialectica would not suffice.

\begin{defi}[{\bf Contraction restriction $\maltese$}]\label{DefContR}%
We upgrade the $\bigstar$ restriction (cf. Remark~\ref{RmkContract}) on the
{\em computationally relevant contractions} (those over refutation relevant open
assumptions $A$), such that the interpretation $\ldIntB{A}$ must be decidable
(rather than strictly quantifier-free). %
This applies to contexts $\spDelta$ of $\IndNatL$ as well, cf. Section~\refp{SeCIndNat}
\end{defi}
In the new modal context one needs to take into account also the translation of the
necessity operator, as this introduces new quantifiers. These may alter the decidability
of the translated formula (relative to the corresponding non-modal formula obtained
by wiping out all instances of $\necsy$)\mssp.
\begin{exas}\label{exas}
Let \lbmh{T(x,y,z)} be a decidable predicate such that
\lbmh{H(x,y) \tspc\EqBDL\tspc \ex z \nspc T(x,y,z)} is not decidable\footnote{\mssp
\Eeg take Kleene's $T$ predicate which is expressible in Peano Arithmetic,
hence also in $\VSys$, so that $H$ expresses the Halting Problem
\QT{program with code $x$ halts on input $y$}.}.
Then \lbmh{P(x) \tspc\EqBDL\tspc \Bfa y \tspc \Bfa z \nspc \lnot \tspc T(x,y,z)}
can be a contraction formula, whereas
\lbmh{P^{\necsy}(x) \,\EqBDL\,
\Bfa y \tspc \necsy \tspc \Bfa z \; \lnot \tspc T(x,y,z)}
cannot, as its translation is
\lbmh{\Bfa z \nspc \lnot \tspc T(x,y,z)},
an undecidable formula, since
\Array{rcl}{\VSys \tspc &\prfg& \tspc \ldInt{P^{\necsy}(x)}{}{y}
\nspc \; \dtLoQuiv \; \nspc \dtnot \tspc H(x,y)}

On the other hand, both
\bmh{\Bfa z \nspc (3 z \neq x) \dtand \!\Bfa y \nspc ( 2 y \neq x)} and
\bmh{\Bfa z \nspc (3 z \neq x) \dtand \!\necsy \Bfa y \nspc (2 y \neq x)}
can be contraction formulas, as \bmh{\Bfa y \nspc (2 y \neq x)} is decidable.
\end{exas}
Thus, given that there is no generic algorithm for the decidability of first-order
formulas over \bmc{\NN} the user needs to supply a boolean term and a proof that
the respective term is equivalent to the translation of the contraction formula.
\Eeg add \lbmh{\Bfa y \nspc (2 y \neq x) \tspc\loquiv\tspc \atom{\!\Odd{x}\!}}
as \emph{global assumption} (cf. \cite{MinLogDoc}), see also Footnote~\ref{FntDF}\mssp.

\section{Modal and light modal functional interpretations}
\label{SeCLMD}%
We prove below that Statement \lref{soundness} ({\em generic soundness})
is valid for parameter instances \mtparam{\MSys}{\VSys}
(modal Dialectica) and \mtparam{\MSysL}{\VSys} (light modal Dialectica),
which share the same \lbmp{\metaVSys \equiv \VSys}
Recall from Definition~\ref{DfNecIntro} of \mssp\InecsY \mssp that the %
restriction on the premise sequent is that
\lbmh{\ux_0 \NotIn \cup_{i=1}^n \tspc \FV{\ut_i}} in its
(light) modal functional translation
\[ \avar{a_1}{\ldInt{A_1}{\ux_1}{\ut_1}}, \ldots,
\avar{a_n}{\ldInt{A_n}{\ux_n}{\ut_n}} \prfg \ldInt{A_0}{\ut_0}{\ux_0}\,.\]%
This ensures that the introduction rule $\faIB$ can be applied
for variables \mbox{$\ux_0$} and thus the conclusion sequent
\bmh{\,\avar{a_1}{A_1}, \ldots, \avar{a_n}{A_n}\,\mprfg\,\necsy A_0\,}
is witnessed by exactly the same realizers as those constructed for
the premise sequent \Lbmh{\spGamma \mprfg A_0}.

\begin{lem}[interpretation of $S_4$ modal axioms]\label{lem_iS4ma}
Axioms $\AxT$, $\AxTc$, $\AxF$, $\AxFc$ and $\AxK$ are realizable
in $\VSys$ under the (light) modal Dialectica translation.
\end{lem}
\begin{proof}
The translation of $\AxT$ is $\ldInt{\necsy A \dtimp A}{\ug}{\ux, \uy}
\EquIVll \Bfa \uv \ldInt{A}{\ux}{\uv} \tspc \dtimp \tspc \ldInt{A}{\ug \ux}{\uy}$
and we can take $\ug$ to be the identity $\lbda{\ux}{\ux}$. Similarly,
the translation of $\AxTc$ is
\[
\ldInt{A \dtimp \wpossy A}{\uf}{\ux,\uy}
\EquIVll \ldInt{A}{\ux}{\uf \ux \uy} \dtimp \ex \uu \ldInt{A}{\uu}{\uy}
\]
and we can take $\uf$ to be the projection $\lbda{\ux \uy}{\uy}$.
For $\tspc\AxF\tspc$ and $\tspc\AxFc\tspc$ it is immediate that
$\ldIntB{\necsy A} \equiv \ldIntB{\necsy \necsy A}$
and also
$\ldIntB{\wpossy A} \equiv \ldIntB{\wpossy \wpossy A}$,
thus the realizer is again the identity in both cases.
In the translation of $\tspc\AxK\tspc$ below, we take
$\uU \EqBD \lbda{\uf, \ug, \ux}{\ug \ux}$, which can easily be
proved to be a realizer.
\Array{@{\hspace{-2pt}}r@{\;}c@{\;}l}{\ldInt{\AxK}{\uU}{\uf,\ug,\uxp}&\equiv&
  \cLDint{\necsy(A \dtimp B)\ \dtand\ \necsy A}{\uf,\ug,\uxp}{}
\ \dtimp\ \ldInt{\necsy B}{\uU(\uf,\ug,\uxp)}{}\quad \equiv\\[8pt]
&\equiv&
\fa \ux,\uv \, ( \ldInt{A}{\ux}{\uf \ux \uv}\dtimp
\ldInt{B}{\ug \ux}{\uv} )\ \dtand\ \fa \uy \ldInt{A}{\uxp}{\uy}
\ \dtimp\ \fa \uvp \ldInt{B}{\uU(\uf,\ug,\uxp)}{\uvp}}
\end{proof} %
\begin{spacing}{1.2}
From Lemma \lref{lem_iS4ma} and the comment above it, we obtain {\it soundness of modal Dialectica}
as Statement~\ref{soundness} \mtparam{\MSys}{\VSys} and {\it soundness of light modal Dialectica}
as Statement~\ref{soundness} \mtparam{\MSysL}{\VSys}.
The next result pictures the actual limits of our modal adaptation of \Goedel's functional interpretation.
\end{spacing}%

\begin{thm}[\Tunrealizability of $S_5$ defining axiom]\label{ThmTUnrealS5}
{\em Axiom $\AxV: \wpossy A \dtimp \necsy\wpossy A$ is not realizable (in general) under
the (light) modal Dialectica translation (by primitive recursive functionals of finite type).}
\end{thm}
\begin{proof}
The translation of $\AxV$ is a formula of shape $B(\uz) \dtimp \Bfa \uz B(\uz)$ for which we would need to construct terms $\utA\in\Term$ so that $B(\utA) \dtimp \Bfa \uz B(\uz)$ is (classically) valid\footnote{\mssp The statement of existence of a (light) modal Dialectica realizer for $\AxV$ amounts to the Drinker's Paradox, a showcase example for a non-constructive principle (made popular by Smullyan in pp. 209--211 of \cite{smullyan}--14C--250 and taken by Barendregt in the context of computer-assisted proofs, cf. \cite{barendregt1996quest}--Section~4.5, pp.~54--55). It should therefore be unsurprising that $\AxV$ is not generally realizable by an interpretation of computational nature.}. We assume $\uz$ is not empty (or else \lbmh{\AxV} required no realizer at all) and note that Statement~\ref{soundness} forces $\uz\NotIn\FV{\utA}$. Marginally, any such type-corresponding terms are good for the case when $\Bfa \uz B(\uz)$, \ie \bmc{\Bfa \uz \exists \ux \ldInt{A}{\ux}{\uz \ux}} holds (in Peano Arithmetic $\PAom$). %
Whenever $B(\uz)$ amounts to a predicate falsified for a set of values corresponding to $\uz$, any such constructible inhabitants would realize $\AxV$ by invalidating the premise of its translation (\eg for \bmc{A \equiv \Bfa z (z =_{\NN} \Zero)} \bmc{B(z) \equiv z =_{\NN} \Zero} with any non-zero number a realizer).

Many instances of $\AxV$ are nonetheless unrealizable, like whenever $A$ is a universal formula whose negation cannot be witnessed constructively. For example, take $A \tspc \EqBD \Bfa z \dtnot T(x, y, z)$ with Kleene's $T$ predicate: $\AxV$ then translates to $\dtnot T(x, y, z) \dtimp \Bfa z \dtnot T(x, y, z)$, equivalent to $H(x, y) \dtimp T(x, y, z)$. A realizer $\tA\lbrack x, y\rbrack$ for $z$ cannot be expressed in \bmc{\Term} as that would imply such an Universal Turing Machine (UTM) existed, while the mere existence of a total UTM enfolds decidability of the Halting Problem \bmh{H} (cf. Examples~\ref{exas}).%
\end{proof}

Notice that $ \wpossy \ex \! \ux A $ is akin to Berger's uniform existence
$ \{ \exists \ux\! \}\tspc A $ from \cite{Berger(93)}, where one does not care about
the witness for $ \exists \ux $ (which is actually deleted from the extraction).
We can thus see $\wpossy$ as an extension of Berger's appliance to more general formulas
than just existential ones.

On the other hand there are situations when $\necsy$ and
$\wpossy$ are too general contrivances and separate annotations for each quantifier are
a better answer for the problem at hand. In some of these cases it may still be possible
to use the modal operators if one changes the input specification and its proof.

\begin{defi}[necessary formulas]\label{DfNecsyFmla}
Formulas $A$ such that $\tspc \mprfg \tspc A \dtimp \necsy A \tspc$ (is provable)\mssp.
\end{defi}
\noindent
Also due to $\AxT$, it follows that
$\nspc \mprfg \tspc A \, \dtLoQuiv \, \necsy A \,$ for
any necessary formula: placing $\,\necsy\,$
in front of such $A$ would be logically redundant (this is akin to Prawitz's
\QT{essentially modal} formulas in \cite{Prawitz(65)}VI.$\S2$, `second version',
see Section~2 of \cite{MartinsNDML} for a concurrent approach).

We say that an occurrence of $\,\necsy\,$ is {\em meaningful} (i.e., non-redundant) in front
of any formula that is not necessary cf. Definition~\refp{DfNecsyFmla}

Note that all refutation irrelevant formulas are necessary formulas. It is
easy to see that some of the refutation relevant formulas are necessary, e.g.,
$\Bfa \ux \LogFalsity$ and $\Bfa \ux \LogTruth$ (in fact any $A$ s.t. $\nspc \mprfg A \nspc$
or $\nspc \mprfg \tspc \dtnot A \nspc$ in $\nspc \MSys$ or $\nspc \MSysL$).
However, even if such formulas syntactically
do require challengers, these functionals turn out to be redundant and can soundly
be discarded by a $\necsy$, without the need to change any other component of the
input proof. In fact, {\em a formula $A$ is necessary iff it can be proved equivalent
(in $\nspc \MSys$ or $\nspc \MSysL$) to a refutation irrelevant formula $B$}.
Indeed, for a necessary $\tspc A \tspc$ take $\tspc B \nspc \EqBDL \tspc \necsy A \tspc$.
For the converse we can use the long implication
$\tspc A \dtimp B \dtimp \necsy B \dtimp \necsy A \tspc$, where for the last
implication a contextless \,\InecsY\, together with $\tspc \AxK \tspc$ was used.
\dtLBrack see also \cite{Prawitz(65)}VI.$\S2$ for {\em modally closed}
formulas\dtRBrack

Therefore, the \SQT{necessary} class captures those formulas whose
negative  computational content can always be erased regardless of the context in
which they are used. On the other hand, there are cases when $\necsy$ can soundly
be applied to a non-necessary formula, leading to cleaner (and thus better) extracted
programs (see Section~\ref{SeCEx} below).

\begin{rem}[non-standard modal]\label{Rmkmvspsm}
It would appear that our Arithmetic $\MSys$ is able to prove new modal
theorems and even sentences that are invalid in \Schuette's semantics.
On the other hand, our $\maltese$ restriction is not present in the usual
first-order modal logic systems, thus some of the classical modal theorems
will no longer be theorems of $\MSys$.

Yet we suspect we are not far from Prawitz's VI.$\S4$ `fourth version' for
$\CSpFive$ with {\it discharge function} for normalization.

The Barcan formula \lbmh{\Bfa z \necsy A(z) \dtimp \necsy \Bfa z \tspc A(z)}
is inadmissible in our modal systems (it is \Tunrealizable in general,
similar to $\AxV$); although invalid in \Schuettes $S^{\star}_{4}$
(cf. Anmerkung at the end of \cite{Schuette}.I.$\S3$), it is provable in Prawitz's
$\PcsFive$ for modally closed $A$ (see page 78 of \cite{Prawitz(65)}VI.$\S2$).
However, the Converse Barcan formula \lbmh{\necsy \Bfa z \tspc A(z) \dtimp \Bfa z \necsy A(z)} is admissible (it is bluntly realizable, similar to $\AxT$). We thus suspect that some form of an increasing domain semantics will be suitable for our systems;
see Sections~2.5, 2.9 of \cite{BraunerFOML}.
\end{rem}
\subsection{Modal induction rule}\label{SecModIndRule}
As first argued in \cite{HernestOliva}, induction (for numbers, but more
generally also for lists, as algebra $\NN$ is a particular case of inductively
defined lists) should rather be treated in a Modified Realizability style whenever
possible under Dialectica extraction. In our non-standard modal context we can
introduce the following {\em modal  induction} rule for %
$\MSys$ and $\MSysL$, which is defined with a Kreisel implication at the step:
\begin{center}
\begin{tabular}{|@{\qquad}c@{\qquad}|}\hline\\[-6pt]
\pt \spGamma\ \prfg\ \necsy A(\Zero) \qquad
\necsy \spDelta \ \prfg\ \necsy A(n) \; \dtimp \;A(\Succ n) \jst
\spGamma \, , \, \necsy \spDelta \ \prfg \ \necsy A(n) \usg{\ \IndNatM} \ept\\[14pt]
\hline\end{tabular}\\[8pt]
\end{center}
This is an upgrade of the similar rule from \cite{HernestOliva} (given at the
linear logic level, see also \cite{Oliva[HFILIL]}), as it allows for non-empty
contexts. While the base context $\spGamma$ is unrestricted, the step context
$\necsy \Delta$ is made entirely of refutation irrelevant assumptions of shape
$\necsy D$.

Thus the step context restriction as for $\modIndNat$ is satisfied by default, since it only concerned refutation relevant assumptions\mssp\footnote{\mssp The decidability of their translations in $\VSys$ were needed for case distinction in their corresponding challenge realizers, cf. Section~\ref{SeCIndNat} for \bmc{\IndNatL} which is the same for \bmc{\modIndNat} only with term-equivalent $\ldIntB{B}$ by default provided by the user at \ptref{EqIndOne}\mssp.}\mssp.
Note that if $D$ already is
refutation irrelevant, placing $\necsy$ in front of $D$ is somewhat redundant.
We could refine $\IndNatM$ by splitting the step context into $\tspc \prm{\Delta} \tspc$
which consists of refutation irrelevant assumptions not of shape $\necsy D \tspc$
and $\tspc \scd{\Delta} \EquiVl \necsy \Delta \tspc$. Nonetheless such
$\tspc \prm{\Delta} \tspc$ would only contain necessary formulas
(cf. Definition~\ref{DfNecsyFmla}).

The treatment of $\IndNatM$ under (light) modal Dialectica is much easier than
the one of $\modIndNat$. In fact $\IndNatM$ is a good simplification of $\modIndNat$
for situations when the whole context is made entirely of refutation irrelevant
assumptions but $\tspc A(n) \tspc$ is a refutation relevant formula. The challenger
for $\tspc A(n) \tspc$ in the step conclusion would be unneededly produced during the
treatment of such $\modIndNat$, as it becomes no part of any of the witnesses
for the conclusion sequent. Placing $\necsN$ in front of the negatively
positioned $\tspc A(n) \tspc$ thus ensures a minimal optimization brought by
$\IndNatM$, in this particular case simply by elimination of redundancy:
the conclusion witnessing terms are the same as for $\tspc\IndNatL$
(\mssp cf. Section~\ref{sec_motiv}\mssp)\mssp.

A more serious optimization concerns the challengers of $\ldIntB{C}$ for refutation
relevant assumptions $C$ from the $\spGamma$ context. These are simply preserved
by $\IndNatM$, while under $\modIndNat$ they would include the challengers for the
step $\tspc A(n) \tspc$. If $\tspc A(n) \tspc$ were refutation irrelevant,
it would still make sense to use $\IndNatM$ instead of $\modIndNat$, if one
is not interested in the challengers for the refutation relevant assumptions
from the step context.

While for such particular instances of $\modIndNat$ we already have
the preservation of challengers for refutation relevant assumptions strictly
from $\spGamma$, still challengers for the refutation relevant
step assumptions are more complex in the conclusion sequent (they include
a meaningful \Goedel recursion, even though here a challenger for the step
negative $\tspc A(n) \tspc$ is no longer comprised since it does not exist).
Thus $\IndNatM$ can bring an improvement over $\modIndNat$ by wiping out
the step challengers altogether, should these not be needed in the global
construction of the topmost realizers for the goal specification.

It turns out that $\IndNatM$ strictly optimizes $\modIndNat$ in many (if not most) situations.
Yet $\modIndNat$ will be employed whenever $\IndNatM$ simply cannot be applied for the goal at hand.\\[-24pt]
\subsubsection*{Modal induction rule --- technical details}
\noindent
We are given both the following
\begin{eqnarray}\label{MEqIndBase}
\ldInt{\spGamma}{\uu}{\ugm} & \prfg & \fa \uy \, \ldInt{A(\Zero)}{\ur}{\uy}\\[6pt]
\nonumber %
\ldInt{\necsy\spDelta}{\uz}{}&\prfg&
\fa\uyp \, \ldInt{ A(n) }{ \ux }{ \uyp } \ \dtimp
\ \ldInt{ A(\Succ n) }{ \us \ux }{\uv}
\end{eqnarray}%
Since \Lbmh{\uv \, \NotIn \, \FV{ \ldInt{ \necsy \spDelta }{ \uz }{}} }
and \Lbmh{\uv \, \NotIn \, \FV{\tspc \Bfa \uyp \, \ldInt{A(n)}{\ux}{\uyp} } }
from the latter we easily obtain%
\begin{eqnarray}\label{MEqIndStep}
\ldInt{\necsy\spDelta}{\uz}{}&\prfg&
\fa \uyp \, \ldInt{A(n)}{\ux}{\uyp}\ \dtimp\
\fa \uv \, \ldInt{A(\Succ n)}{\us \ux}{\uv}
\end{eqnarray}
\begin{spacing}{1.2}%
With
\Lbmh{ t\tspc [ \tspc n \tspc ] \ \EqBDL \ \Rec \; n \; r \; ( \lbda{n}{s} ) }
for every corresponding pair \lbmh{\pair{r\in\ur}{s\in\us}}
we show by induction on \lbmh{n} in \lbmh{\VSys} with
base context \lbmh{ \InLdInt{\spGamma}{\uu}{\ugm} }
and step context
\lbmh{ \InLdInt{\necsy\spDelta}{\uz}{} } that
\end{spacing}%
\begin{eqnarray*}
\ldInt{\spGamma}{\uu}{\ugm}\,\ ,\,\ \ldInt{\necsy \spDelta}{\uz}{}
&\prfg&\fa\uv\,\ldInt{A(n)}{\ut[n]}{\uv}
\end{eqnarray*}
\begin{spacing}{1.2}%
As \lbmh{ \ut[\Zero] \, \EquiVl \, \ur } the base is given by \ptref{MEqIndBase}
and the step follows from \ptref{MEqIndStep} with
\lbmh{ \ux \tspc \mapsto \tspc \ut[n] }
since $\ut[\Succ n] \, \EquiVl \, \us \,\ut[n] \tspc$. Thus challengers
\lbmh{\ugm} are simply preserved for \lbmh{ \ldIntB{\spGamma} } and witnesses
\lbmh{\ut[n]} are easily constructed for \lbmh{ \ldIntB{ \necsy {A(n)} \tspc} }
in the conclusion sequent of \bmp{\IndNatM}
\end{spacing}%

\begin{rem}
Our modal induction rule is equivalent to a special case of $\IndNat$, since a
$\necsy$ can be placed in front of $A(\!\Succ n)$ from the step sequent
of $\IndNatM$. The equivalence of the two formulations for the step sequent
can easily be proved using $\AxT$\mssp, $\AxF$\mssp, $\AxK \tspc$ and
$\tspc \mINcsy$. Extracted terms are the same and the verifying
proof only gets more direct.
\end{rem}
\subsection{Revisited examples}\label{SeCEx}
The weak extensionality of modal input systems $\MSys$ and $\MSysL$ can
be expressed by means of the following {\em modal compatibility axiom}
(the usual compatibility axiom, but with the outward implication changed
to a Kreisel implication; see \cite{Oliva[HFILIL]}--Introduction
for the akin formulation in linear logic using a `Kreisel modality' $\kNecsy$)
$$ \AxCompatM: \qquad \necsy \tspc ( \ux =_{\uro} \uy) \quad
\dtimp \quad  B(\ux) \quad \dtimp \quad B(\uy) $$
By straightforward calculations, it is easy to see that $\AxCompatM$ is realizable
under (light) modal Dialectica by simple projection functionals, with the verification
in the fully extensional $\VSys$ given by the corresponding compatibility axiom
\bmp{\AxCompat} The realizing terms are same \lbmh{f \dtcsp g} as for
\lbmh{\RlCompat_{\dtRho}} at the end of Section~\refc{SeCExtens} here
just grouped in tuples.

In \cite{HernestOliva} the following class of examples was considered:
theorems of the form
\begin{equation} \label{form1}
	\Bfa \ux \,A \ \ \dtimp \ \ \Bfa \uy \,B \ \ \dtimp \ \ \Bfa \uz \,C
\end{equation}
possibly with parameters, where the negative information on $\ux$ is
irrelevant, while the one on $\uy$ is of our interest. Then it must be possible
to adapt the proof of \ptref{form1} to a proof in \lbmh{\MSys} or \lbmh{\MSysL}
of \Lbmp{(\necsy \Bfa \ux \tspc A \tspc) \; \dtimp \; \Bfa \uy \tspc B
\; \dtimp \; \Bfa \uz \tspc C \tspc}
As noticed by Oliva in \cite{Oliva[HFILIL]}, the Fibonacci example first
treated with Dialectica in \cite{Hernest[DCM06]} falls into this category.
Oliva also suggested an interesting example, which motivated the definition of
our positively computational quantifier $\fapl$ (\mssp cf. Example~\ref{DfLDT}
and Definition~\ref{DfLMDI}\mssp):
``Any infinite decidable set $P$ of natural numbers contains elements which are arbitrarily far apart''.
The claim can be formalized (in an extension of \lbmh{\VSys} with proper
predicate symbols) as follows:
\begin{eqnarray*}
  \fa x\, \ex y\; \big( \tspc y > x \,\dtand\, P(y) \tspc \big)
&\dtimp& \fa d\; \ex \tspc n_1 \dtcsp n_2 \;
\big( \tspc n_2 \,>\, n_1 + d \;\dtand\; P(n_1)\; \dtand\; P(n_2) \tspc \big)
\end{eqnarray*}
This statement can be proved only via a contraction on the premise, and as a result
(the negative universally quantified) \lbmh{x} gets refuted by a term involving
case distinction on \lbmp{\ldIntB P}

If nonetheless only the witnesses of \lbmh{n_1}
and \lbmh{n_2} are needed, then the redundant challenge for \lbmh{x} can simply
be discarded by means of a $\necsy$ in front of the premise, effectively
applying a Kreisel implication. This example is of the form \ptref{form1} and
was extensively treated in Section~4 of \pCite{HerTrifonLDR} It can even be
treated with the hybrid Dialectica from \scCite{HernestOliva} we here only
bring the more instrumental solution.

The example can be extended so that the premise becomes more involved
(cf. \cite{TriffonPhD}, Example 5.3 on page 114):
\begin{equation} \label{EqInvEx} \fa m \;
\big(\tspc \ex n \; Q(n,m) \;\dtimp\; \ex n_1 \; Q(n_1, \Succ m)
\tspc \big) \  \,\dtimp\, \  \big( \tspc \ex n_0 \; Q(n_0, \Zero)
\  \dtimp \  \ex n_2 \; Q(n_2, \Succ \Succ \Zero) \tspc \big)
\end{equation}
Again, a contraction must be used, and two semi-computational quantifiers
need to be applied in order to erase the negative computational content.
The light specification corresponding to \ptref{EqInvEx} would then be written as:
\begin{eqnarray*}
\fapl m \; \big( \tspc \expl n \; Q(n,m) \; \dtimp \;
\exf n_1 \; Q(n_1, \Succ m) \tspc \big)
&\  \dtimp \  &\exf n_0 \; Q(n_0, \Zero)
\  \dtimp \  \exf n_2 \; Q(n_2, \Succ \Succ \Zero)
\end{eqnarray*}
This solution is withal not desirable, as the light annotations would only
apply to a special class of binary relations $Q$ for which the witness $n_1$
for \lbmh{Q(n_1, \Succ m)} does not depend computationally on the witness $n$ for
\lbmh{Q(n,m)} for any \bmc{m} hence reducing the generality of the claim.
A fix would then be to extend the light annotations to implications, as in \cite{TriffonPhD}.

However, a much simpler and more elegant approach is to use a Kreisel implication,
by placing $\necsy$ in front of $\Bfa m \;
\big(\tspc \ex n \, Q(n,m) \,\dtimp\, \ex n_1 \, Q(n_1, \Succ m)
\tspc \big)$ at \reffp{EqInvEx}  The negative content of the main premise
will thus be fully erased and the positive one will be fully preserved,
achieving a Modified Realizability effect.
We also mention a proof for the `integer root example' (first considered in \cite{Berger(95)}):
``every unbounded integer function has an integer root function''.
The statement can be formalized (in negative arithmetics) as follows:
\begin{equation}\label{EqKreis}
  \Bfa x \, \ex y \; \big( \tspc f(y) \tspc > \tspc x \tspc \big) \
\,\dtimp\, \ \Bfa m \; \Big(\nspc f( \Zero ) \leq m \; \dtimp \; \ex n \,
\big(\tspc \tspc f(n) \tspc \leq \tspc m \tspc < \tspc f(\Succ n) \tspc \big) \nspc \Big)
\end{equation}

The claim can be proved by contradiction using $n$-induction for the formula
\lbmp{f(n) \tspc \leq \tspc m} \mssp In addition to computing the integer
root, \Goedel's Dialectica also extracts a complex recursive counterexample
for \bmc{x} with a case distinction on each step (cf. \cmCite{TriffonPhD} section 3.2).
This term challenges the outermost premise
\lbmh{\Bfa x \, \ex y \; ( \tspc f(y) \tspc > \tspc x \tspc )}
which actually constitutes the refutation relevant context shared by both
the base and the step formulas of the induction.

The undesired negative content can be erased by `Kreisel-izing' the outermost
implication of \reffc{EqKreis} thus converting the context to a necessary one,
hence allowing for the application of the modal induction rule. As a result,
only the integer root gets synthesized (\mssp the realizer for \lbmh{n} as
function of \lbmh{m}) and additional artifacts are omitted.

Note that, in contrast to the previous two examples, this proof is intrinsically classical,
so Modified Realizability alone is not applicable in this case. Using
\lbmh{\fapl x} would nevertheless still achieve the same cleaning effect
(cf. \cmCite{TriffonPhD} section 5.6.1)\mssp.

\subsection{Proof that \texorpdfstring{$\necsy$}{Box} is a strict addition to the light system}\label{SecBoxNoSS}\begin{spacing}{1.2}
  The (modal) translation of an input schemata
\lbmh{\big( \fa n \ex m \nspc A(m,n) \dtimp \fa n \ex m \nspc B(m,n)\nspc\big) \tspc \kimp \tspc \dtnot \fa k\tspc C(k)}
with decidable predicates $A,B,C$ is
\lbmc{\fa h,n \,\lbrack\tspc A ( h (g h n) , \tspc g h n ) \tspc \dtimp \tspc B (f h n , n) \tspc \rbrack \tspc \dtimp \tspc \dtnot C( K f g )}
where \lbmh{K} is the witness variable and \lbmh{f,g} are challenge variables.

Such specification cannot be produced by means of light quantifier decorations of the schemata
\lbmp{\big( \fa n \ex m \nspc A(m,n) \dtimp \fa n \ex m \nspc B(m,n)\nspc\big) \dtimp \dtnot \fa k\tspc C(k)}

Below is the small \Minlog program that was used to carry out the modal translation; the raw \Minlog output
has been processed for readability. \dtLBrack \texttt{@@} binds a pair of types \dtRBrack\end{spacing}{\footnotesize
\begin{verbatim}
(load "C:\\minlog\\initDan.scm") ; initial system load, adapted to Windows pathnames
(load "C:\\minlog\\etsmdA.scm") ; library for modal Dialectica that adapts src/etsd.scm
(libload "nat.scm") ; library for numbers that also defines n, m, k of type `nat'
(add-predconst-name "A" "B" (make-arity (py "nat") (py "nat")))
(add-predconst-name "C" (make-arity (py "nat"))) ; no computational vars for predconsts
;; (add-var-name "f" "g" (py "((nat=>nat)=>nat=>nat)"))
;; (add-var-name "h" (py "nat=>nat")) ; below `F' is Minlog's decidable falsum
(define oG (pf "(all n ex m  A m n -> all n ex m B m n) --> (all k C k -> F) "))
(define mdoG (formula-to-md-formula oG)) ; (pretty-print mdoG)
; (add-var-name "K" (py "(((nat=>nat)=>nat=>nat)@@((nat=>nat)=>nat=>nat)=>nat)"))
;;; ex K all f,g { all h,n [ A(h (g h n) , g h n) -> B(f h n, n) ]  -> C (K f g) -> F }
\end{verbatim} }

\subsection{\texorpdfstring{Illustrative example: finitary Infinite Pigeonhole Principle (cf. \cite{RatiuTrifonov})}{Illustrative example: finitary Infinite Pigeonhole Principle}}
In his PhD thesis (cf. Chapter~5 of \cmCite{TriffonPhD} in particular Section~5.6.2) the second author explains that, under the light Dialectica of \cite{Hernest[PhD]}\mssp\footnote{\mssp The second author's adaptation of the first author's archived code in \cite{MinLogDoc} is a structural permutation of equivalent complexity. It lacks the semi-computational quantifiers, considered for a future upgrade of \cite{MLFD}.}, three uniform quantifiers need to be inserted in order to remove the negative computational content from three universally quantified formulas inside the proof\mssp\footnote{%
  Note that the term in Figure~5.3 of \cite{TriffonPhD} is a hand-compiled version of the expression of Table~5.3. The term and the expression denote one and the same program, but in Table~5.3 the extraction of the program is shown in a stepwise manner, so that every step can be related to the proof and to the interpretation. Figure~5.3 represents an operationally cleaner Scheme program. No normalization is happening between Table~5.3 and Figure~5.3: the second author avoided it, as (uncontrolled) normalization can produce a slower program.}.
It turns out that this can be achieved by inserting a single $\necsy$ in the formulation of the corollary he is proving (Unbounded Pigeonhole Principle)\mssp\footnote{\mssp In front of the conjunction \lbmc{\mathrm{Decr}(l,n)\cLand \mathrm{Same}(l,n)} see Corollary~3.6 on page 63 of \cite{TriffonPhD}\mssp. At the time of writing of \cite{TriffonPhD} the \Minlog implementation of $\fanc$ was not operational for proofs involving case distinction (for numbers) like the one produced by the second author for comparison with the A-translation approach (cf. \cite{MinLogDoc}--14.1, \cite{pcbook}--7.3). To address this problem, the first author rearranged the input specification in \cite{MLFD} so that two $\dtimp$ can be rewritten as $\kimp$, otherwise the modal input proof essentially is equivalent to the proof used by the second author in \cite{TriffonPhD}\mssp. The case distinction treatment of $\fanc$ was subsequently fixed in \Minlog and thus any of the two versions of the proof (modal, or light-only) may now be used.}\mssp.

The treatment of the example now becomes simpler, with the same synthesized term as the one displayed by the second author in his thesis. The advantage of modal Dialectica is that in the input proof one only needs to check the uniformity condition once for the $\necsy$ (\mssp logically pushed in front of \lbmh{\mathrm{Decr}(l,n)\cLand \mathrm{Col}(l,n)} from the intermediate lemma\mssp) rather than two times for $\fanc$ introductions. The paradigm here is that one can outline the optimizations ``en masse'' rather than piece by piece.

Note that the program (manually) extracted by the second author basically is the same as the one described by Kohlenbach in Section~11.4 of \cite{KohlenBook} by means of Oliva's finite bar recursion, cf. Section~2.1 of \cite{OlivaFBR}, see also \cite{GaspaKohlTao}. The first author carried out the implementation in \Minlog by means of the Kreisel implication and automatically obtained the bettered Scheme program from Figure~5.3 of \cite{TriffonPhD}, see the Appendix sections in \pCite{HTA}

\section{Conclusion and future directions}\label{SeCFut}

Modal Dialectica provides the means of using both Modified Realizability and
\Goedel's Dialectica at the same time for more efficient program synthesis. %
This was already the case for the hybrid Dialectica of \cite{HernestOliva},
but here we avoid the detour to the linear logic substructure. Disregarding
the light quantifiers, modal Dialectica represents (directly at the
supra-linear level) a good combination of the original proof interpretations,
with the possibility of carrying out both in a sound way on certain input proofs,
insofar as some implications of the input specification can be `Kreisel-ized'.
At the extreme, Modified Realizability is obtained from Dialectica, see also
the comments above Definition~\ref{DfMDI}\mssp. \Eeg
\begin{eqnarray}\nonumber
\ldInt{(A \kimp B) \kimp C}{\uH}{\ug,\up}
&\equiv&
\fa \ux,\uv \ldInt{A \kimp B}{\ug}{\ux,\uv}
\dtimp \ldInt{C}{\uH \ug}{\up}\\
&\equiv&
\fa \ux,\uv \ \big(\fa\uy \ldInt{A}{\ux}{\uy} \dtimp \ldInt{B}{\ug\ux}{\uv}\big)
\dtimp \ldInt{C}{\uH \ug}{\up}
\end{eqnarray}
Why not invoke a Modified Realizability (MR) extraction procedure for $B\dtimp C$ instead of processing $B\kimp C$\,?
Per se, MR requires strong existential quantification;
even in combination with (refined) A-translation (cf. \cite{Berger(02)}),
restrictions are in place for the shape of the goal formula.
Thus it is modal Dialectica that provides the fully modular approach. \Eeg the Dialectica
extracted term from the (classical) proof of IPP (Infinite Pigeonhole Principle) can be
(re)used further in the synthesis of programs that employ IPP as lemma (such as the
Unbounded Pigeonhole Principle).

\begin{spacing}{1.2}
A natural continuation of the work reported in this paper concerns the addition
to our input systems of strong (intuitionistic) elements. Besides the strong $\iex$
and its light associated $\iexnc$ (originally from \cite{Hernest[PhD]} where it
was denoted $\ol{\iex}$, see also \cite{TriffonPhD}\mssp), {\em strong possibility}
$\possy$ also needs to be considered as the intuitionistic dual of necessity $\necsy$.
\end{spacing}

\begin{spacing}{1.4}
The following clauses would then be added to Definition~\ref{DfMDI} for
getting the {\em strong modal Dialectica} interpretation
$ \InLdInt{ \iex z \nspc A(z) }{ z, \uf }{ \uy }
\ \EqBD \ \InLdInt{ A(z) }{ \uf }{ \uy } $
and
$ \InLdInt{ \possy A }{}{ \uy } \ \EqBD
\ \iex \! \ux \InLdInt{A}{\ux}{\uy} $,
and further
$ \InLdInt{ \iexnc z \nspc A(z) }{ \ux }{ \uy }
\ \EqBD \ \iex z \nspc \InLdInt{A(z)}{\ux}{\uy} $
to Definition~\ref{DfLMDI}  in order to obtain the
{\em strong light modal Dialectica}  interpretation.
\end{spacing}

Intuitionistic (light) modal arithmetical systems will first be considered
at input for \SQT{strong} program synthesis. Then their enhanced classical
counterparts will be interpreted, modulo some negative translation. Such
systems will soundly extend \lbmh{\MSys} with \lbmh{\possy} and $\iex$,
and \lbmh{\MSysL} also with \bmp{\iexnc} Nevertheless, certain restrictions
may need to be applied on \lbmh{\MSys} \AndSlashOr \lbmh{\MSysL}
before attempting such extensions with intuitionistic elements\mssp\footnote{\mssp See \cite{MartinsNDML} for weak normalization of standard first-order classical S5 (with strong existence and strong possibility) and Chapters~4 and 7 of \cite{AKsimpsonPhD} for an intuitionistic account of intuitionistic modal logic\mssp.}\mssp.

In Section~3.2 of \cite{Oliva[HFILIL]} Oliva suggested labelled contexts
in order to deal with the technical difficulties of having both the Kreisel
and the usual (\Goedel) implications in intuitionistic logic $\ILom$. Our
implementation in \Minlog of $\kimp$ identifies those \QT{Kreisel} assumptions
as the ones discharged at \impnc introduction; they are marked so that no
realizer is extracted for their negative side. In the modal language,
we can say that they are \QT{boxed} by means of $\necsy$, which acts as a
\QT{Kreisel} label. The restriction from Definition~\ref{DfNecIntro} then
has to be checked for the proof of the premise of an \impnc elimination.

It is straightforward that the hybrid system with $\kimp$ is fully expressible
in $\MSys$; the question is whether $\MSys$ could nicely be expressed in a
system with the Kreisel implication as primitive, given that
\lbmp{\ldIntB{\necsy A} \;\loquiv_{\VSys}\; \ldIntB{(A \kimp \bot) \dtimp \bot}}
Perhaps a {\it Kreisel negation} $\notK$ were more suitable, with
\lbmp{\ldIntB{\notK A} \;\loquiv_{\VSys}\; \ldIntB{(A \kimp \bot)}}

The design of the monotone variant of modal Dialectica is under construction, since it has been known for some time that a (heterogeneous) combination of modified realizability and classical Dialectica was successfully used by Leu\c{s}tean for proof mining (\mssp cf. \cite{KohlenBook}\mssp) an exceptional approximation result in metric fixed-point theory (\mssp cf. \cite{LLAppPMnli,LeoNIUCW}\mssp). See also \cite{MDH06-LMDPM} for a synthetic analysis of the impact of the precursor of $\necsy$ into Kohlenbach's advanced framework for Proof Mining\mssp; note that our base logical framework is equivalent to the one used by the proof miners, cf. Section~1.1.11 of \cite{Troelstra(73)}, see also \cite{Luckhardt(73)}. Recent works by Powell \cite{DBLP:journals/lmcs/Powell20} and {\c S}ipo{\c s} \cite{SiposRMIUI} would be suitable for implementation in \cite{MLFD}, as indicated by Kohlenbach\mssp.

Last but not least, the interplay between proofs and programs in our non-standard modal systems
may be suitable for the discovery approach of DreamCoder \cite{ellis2020dreamcoder}. Instead of incrementally building (by intervention of human operators) an information system associating realizers to (admissible) proofs of Lemmata (as building blocks for the semi-automated search of programs from prima facie non-constructive proofs of Theorems) we could then have the machine (re)discover \Minlog and upgrade it to its modal variant\mssp. %

\textit{Our \textnormal{\Minlog} variant and implementation of modal Dialectica may be found at:\\ \url{https://triffon.github.io/mlfd}}
\section*{Acknowledgment}
Our first reading of predicate modal logic was \cite{Schuette},
a rare small and complete (at the time of its writing) presentation of the topic,
recommended by Helmut Schwichtenberg. Thanks to Diana Ra\c{t}iu and Iosif Petrakis for logistic support
and to Paulo Oliva for valuable comments on early drafts of this paper.
We also thank the anonymous reviewers which contributed throughout the making of this paper by pertinent remarks.

\bibliographystyle{alpha}
\bibliography{ModFI}
\end{spacing}
\end{document}